\documentclass[12pt]{article}

\usepackage{a4wide}
\usepackage{amssymb}
\usepackage{amsmath,amsthm}
\usepackage[latin1]{inputenc}
\usepackage{graphicx}
\usepackage{hyperref}
\usepackage{enumerate}
\usepackage{tikz}
\usepackage{tkz-graph}
\usetikzlibrary{calc, positioning, decorations, backgrounds,snakes}
\usetikzlibrary{shapes}
\usepackage{mathrsfs}
\usepackage{verbatim}
\usepackage[ruled, linesnumbered, vlined,resetcount]{algorithm2e}
\usepackage{algorithmicx}
\usepackage{mathtools}
\usepackage{subfigure}
\usepackage{multirow}
\usepackage[normalem]{ulem}
\usepackage{array}
\newcolumntype{C}[1]{>{\centering\let\newline\\\arraybackslash\hspace{0pt}}m{#1}}
\usepackage{nccmath}

\newtheorem{definition}{Definition}

\hypersetup{colorlinks=true, linkcolor=blue, citecolor=blue, urlcolor=blue}

\newcommand{\tuple}[1]{\langle#1\rangle}

\newcommand{\mmid}{\!\mid\!}

\newcommand{\subgraph}[2]{\tuple{#1}_{#2}}
\newcommand{\WeaklyInduced}[2]{\langle #1 \rangle^w_{#2}}
\newcommand{\Gat}[1]{{G_{#1}}}
\newcommand{\Gsat}[1]{{G^+_{#1}}}

\newcommand{\Grat}[1]{{G^{\star}_{#1}}}
\newcommand{\GratInit}[1]{{\tilde{G}^{\star}_{#1}}}

\newcommand{\Gsb}[1]{\WeaklyInduced{S_{#1}}{\Gsat{#1}}}

\newcommand{\set}[1]{\{#1\}}

\newcommand{\dom}{\operatorname{dom}}

\DeclareMathOperator*{\argmax}{arg\,max}

\newcommand{\similarity}{\operatorname{sim}}

\title{Active Re-identification Attacks on Periodically Released 
Dynamic Social Graphs}

\author{Xihui Chen$^1$, Ema K\"epuska$^2$, Sjouke Mauw$^{1,2}$ 
and Yunior Ram\'{i}rez-Cruz$^1$\\ 
{\small $^1$SnT, $^2$CSC, University of Luxembourg}\\ 
{\small 6, av. de la Fonte, L-4364 Esch-sur-Alzette, Luxembourg}\\ 
{\small \{xihui.chen, sjouke.mauw, yunior.ramirez\}\@@uni.lu, 
kepuskaema\@@gmail.com}} 

\begin{document}
\maketitle

\begin{abstract} 
Active re-identification attacks pose a serious threat
to privacy-preserving social graph publication. Active attackers
create fake accounts to build structural patterns
in social graphs which can be used to re-identify
legitimate users on published anonymised graphs, even without additional
background knowledge. 
So far, this type of attacks has only been studied in the scenario
where the inherently dynamic social graph is published once.
In~this paper, we present the first active re-identification attack
in the more realistic scenario where a dynamic social graph
is periodically \mbox{published}.
The new attack leverages tempo-structural patterns 
for strengthening the adversary.
Through a comprehensive set of experiments on real-life and synthetic
dynamic social graphs, we show that our new attack
substantially outperforms the most effective static active attack
in the literature by increasing the success probability of re-identification
by more than two times and efficiency by almost 10 times. 
Moreover, unlike the static attack, our new attack is able to remain 
at the same level of effectiveness and efficiency as the publication 
process advances. We conduct a study on the factors that may thwart 
our new attack, which can help design graph anonymising methods with 
a better balance between privacy and utility. 
\end{abstract}


%===================================================
\section{Introduction}
\label{sec:intro}
%===================================================

Social graphs have proven to be a valuable data source for conducting sociological
studies, market analyses, and other forms of complex data analysis.
This creates a strong incentive for the establishment
of a mutually beneficial relation between analysts and data owners.
For analysts, it is of paramount importance to have access to abundant,
reliable social graph data in order to conduct their studies.
For data owners, making these data available to third parties
opens a number of additional business opportunities, as well as
opportunities for improving their social perception by contributing
to the advancement of research. However, releasing social network data raises
serious privacy concerns, due to the sensitive nature of much of the information
implicitly or explicitly contained in social graphs. Consequently, the data needs
to be properly sanitised before publication.

It has been shown that some forms of sanitisation, e.g. removing users'
identities and personally identifying information from the released data,
a process known as \emph{pseudonymisation}, are insufficient
for protecting sensitive information. This is due to the fact that a majority
of users can still be unambiguously \emph{re-identified}
in the pseudonymised graph by means of simple structural
patterns~\cite{LT08,NS09,BDK11,NKA14}. The re-identification subsequently
facilitates inferring relations between users, group affiliations, etc.
A method allowing a malicious agent, or \emph{adversary}, to re-identify
(a subset of) the users in a sanitised social graph is called
a \emph{re-identification attack}.

A large number of anonymisation methods have been proposed for publishing
sanitised social graphs that effectively resist re-identification attacks.
The largest family of graph anonymisation methods
(e.g.~\cite{LT08,Lu2012,UMGA,Chester2013,Wang2014,Ma2015,Salas2015,Rousseau2017,Casas-Roma2017,ZP2008,ZCO09,Wu2010,MRT18,MRT18b})
follows a common
strategy of editing the vertex and/or edge set of the pseudonymised graph in order
to satisfy some formal privacy properties. These privacy properties
rely on an adversary model, which encodes a number of assumptions
about the adversary capabilities. In the context of social graph publication,
there are two classes of adversaries. On the one hand, \emph{passive} adversaries
depend on publicly available information, in case that it can be obtained
from online resources, public records, etc., without interacting
with the social network before publication.
On the other hand, \emph{active} adversaries interact with the network
before the sanitised dataset is released, in order to force the existence
of structural patterns. Then, when the sanistised graph is published,
they leverage these patterns for conducting the re-identification.
Active adversaries have been shown to be a serious threat
to social graph publication~\cite{BDK11,MRT19}, as they remain plausible
even if no public background knowledge is available.
Active attackers have the capability of inserting fake accounts
in the social network, commonly called \emph{sybil nodes},
and creating connection patterns between these fake accounts
and a set of legitimate users, the \emph{victims}. After the publication
of the sanitised graph, the attacker uses these unique patterns
for re-identifying the~victims.

Social networks are inherently dynamic. Moreover, analysts require datasets
containing dynamic social graphs in order to conduct numerous tasks
such as community evolution analysis~\cite{DTSB19},
link prediction~\cite{LZ11} and link persistence analysis~\cite{PK19},
among others. Despite the need for properly anonymised dynamic social graphs,
the overwhelming majority
of studies on graph anonymisation have focused on the scenario of a social
graph being released only once. The rather small
number of studies on dynamic social graph publication have provided
only a partial understanding of the field, as they have exclusively focused
on passive adversary models. Consequently, the manners in which active
adversaries can profit from a dynamic graph publication scenario remain
unknown. In this paper, we remedy this situation by formulating active
re-identification attacks in the scenario of dynamic social graphs.
We consider a scenario where the underlying dynamic graph
is \mbox{\emph{periodically}}
sampled for snapshots, and sanitised versions of these snapshots
are published. We model an active adversary whose knowledge
consists in \emph{tempo-structural} patterns, instead of exclusively structural
patters as those used by the original (static) active adversary. Moreover,
in our model the adversary knowledge is \emph{incremental}, as it grows
every time a new snapshot is released, and the adversary has the opportunity
to adapt along the publication process. Under the new model, we devise
for the first time a dynamic active re-identification attack
on periodically released dynamic social graphs.
The new attack is more effective than the alternative of executing independent
static attacks on different snapshots. Furthermore, it is also considerably
more efficient than the previous attacks, because it profits from temporal
patterns to accelerate the search procedures in the basis of several
of its components.

\vspace{2mm}\noindent
{\bf Our contributions.} The main contributions of this paper are listed
in what follows:
\begin{itemize}
\item We formulate, for the first time, active re-identification attacks
in the scenario of periodically released dynamic social graphs.
\item Based on the new formulation, we present, to the best of our knowledge,
the first dynamic active re-identification attack
on periodically released dynamic social graphs, which constructs
and leverages tempo-structural patterns for re-identification.
\item We conduct a comprehensive set of experiments on real-life
and synthetic dynamic social graphs, which demonstrate
that the dynamic active attack is more than two times more effective 
than the alternative of repeatedly executing the strongest active attack
reported in the literature for the static scenario~\cite{MRT19}.
\item Our experiments also show that, as the number of published snapshots grows,
the dynamic active attack runs almost 10 times faster than 
the static active attack from~\cite{MRT19}. 
\item We analyse the factors that affect the effectiveness of our new attack. 
The conclusions of this study serve as a starting point for the development
of anonymisation methods for the new scenario.
\end{itemize}

\vspace{2mm}\noindent
{\bf Structure of the paper.}
We discuss the related work in Sect.~\ref{sec:related work},
focusing in similarities and differences with respect to our new proposals.
Then, we describe the periodical graph publication scenario, accounting for active
adversaries, in Sect.~\ref{sec:framework}; and we introduce the new dynamic active
attack in Sect.~\ref{sec:attack}. Finally, our experimental evaluation
is presented in Sect.~\ref{sec:experiments} and we give our conclusions
in Sect.~\ref{sec:conclusion}.

%===================================================
\section{Related Work}
\label{sec:related work}
%===================================================

Re-identification attacks are a relevant threat for privacy-preserving 
social graph publication methods that preserve a mapping between 
the real users and a set of pseudonymised nodes in the sanitised release~\cite{LT08,Lu2012,UMGA,Chester2013,Wang2014,Ma2015,Salas2015,Rousseau2017,Casas-Roma2017,ZP2008,ZCO09,Wu2010,MRT18,MRT18b}.

Depending on the manner in which the attacker obtains the knowledge
used for re-identification, these attacks can be divided into two classes:
passive and active attacks. \emph{Passive} adversaries collect
publicly available knowledge, such as public profiles in other social networks,
and searches the sanitised graph for vertices with an exact or similar profile.
For example, Narayanan and Shmatikov~\cite{NS09} used information from Flickr
to re-identify users in a pseudonymised subgraph of Twitter. A considerable
number of passive attacks have been proposed,
e.g.~\cite{NS09,Narayanan2011,Yartseva2013,Pedarsani2013,Nilizadeh2014,Ji2014,Korula2014,Ji2014a}.
On the other hand,
\emph{active} adversaries interact with the real network before publication,
and force the existence of the structural patterns that allow re-identification
after release. The earliest examples of active attacks
are the \emph{walk-based attack} and the \emph{cut-based attack},
introduced by Backstrom \emph{et al.} in~\cite{BDK11}. Both attacks insert
sybil nodes in the network, and create connection patterns between the sybil nodes
that allow their efficient retrieval in the pseudonymised graph.
In both attacks, the connection patterns between
sybil nodes and victims are used as unique fingerprints allowing re-identification
once the sybil subgraph is retrieved. Due to the low resilience
of the walk-based and cut-based attacks, a robust active attack was introduced
by Mauw \emph{et al.} in~\cite{MRT19}. The robust active attack 
introduces noise tolerant
sybil subgraph retrieval and fingerprint mapping, at the cost
of larger computational complexity.
The attack proposed in this paper preserves the noise resiliency of the robust
active attack, but puts a larger emphasis on temporal consistency constraints
for reducing the search space. As a result, for every run
of the re-identification, our attack is comparable
to the original walk-based attack in terms of efficiency
and to the robust active attack in terms of resilience against modifications
in the graph.

Notice that, by itself, the use of connection fingerprints
as adversary knowledge does not make an attack active. The key feature 
of an active attack is the fact that the adversary interacts with the network 
to force the existence of the fingerprints. For example, 
Zou \emph{et al.} \cite{ZCO09} describe an attack that uses as fingerprints
the distances of the victims to a set of hubs. This is a passive attack, 
since hubs exist in the network without intervention of the attacker. 

The attacks discussed so far assume a single release scenario. A smaller
number of works have discussed re-identification in a dynamic scenario.
Some works 
assume an adversary
who can exploit the availability of multiple snapshots, although they only give
a coarse overview of the increased adversary capabilities,
without giving details on attack strategies. Examples of these works
are \cite{TTPC11}, which models a passive adversary that knows the evolution
of the degrees of all vertices; and \cite{ZCO09}, which models another
passive adversary that knows the evolution of a a subgraph around the victims.
An example of a full dynamic de-anonymization method is given in~\cite{DZWG11}.
Although they do not model an active adversary,
the fact that the method relies on the existence of a seed graph makes it
potentially extensible with an active first stage for seed re-identification, 
as done for example in~\cite{Peng2012,PLZW2014}. 
Our attack differs from the methods above
in the fact that it uses an evolving set of sybil nodes that dynamically
interact with the network and adapt to its evolution. 

%===================================================
\section{Periodical Graph Publication in the Presence of Active Adversaries}
\label{sec:framework}
%===================================================

In this section we describe the scenario where the owner of a social network
periodically publishes sanitised snapshots of the underlying dynamic social graph,
accounting for the presence of active adversaries. We describe this scenario
in the form of an attacker-defender game between the data owner and the active
adversary. We first introduce the basic notation and terminology,
and then give an overview of the entire process and a detailed description
of its components.

%---------------------------------
\subsection{Notation and Terminology}
\label{subsec:preliminaries}
%---------------------------------

We represent a dynamic social graph as a sequence
$\mathcal{G}=(G_1, G_2, \ldots, G_i, \ldots)$, where each $G_i$ is a static graph
called the $i$-th \emph{snapshot} of $\mathcal{G}$. Each snapshot
of $\mathcal{G}$ has the form $G_i=(V_i, E_i)$, where $V_i$ is the set of vertices
(also called \emph{nodes} indistinctly throughout the paper)
and $E_i\subseteq V_i\times V_i$ is the set of~edges.

We will use the notations $V_G$ and $E_G$ for the vertex and edge sets
of a graph $G$. In this paper, we assume that graphs are simple and undirected.
That is, $G$ contains no edges of the form $(v, v)$ and, for every pair
$v,w\in V_G$, $(v, w)\in E_G \iff (w, v)\in E_G$.
The \emph{neighbourhood} of a vertex $v$ in a graph $G$ is the set
$N_G(v) =\{w\in V\mid (v, w)\in E\}$, and its \emph{degree}
is $\delta_G(v)=\vert N_G(v)\vert$. For the sake of simplicity,
in the previous notations we drop the subscript when it is clear from the context
and simply write $N(v)$, $\delta(v)$, etc.

For a subset of nodes $S\subseteq V_G$, we use $\subgraph{S}{G}$ to represent
the subgraph of $G$ \emph{induced} by $S$, i.e.
$\subgraph{S}{G}=(S, E_G\cap (S\times S))$.
Similarly, the subgraph of $G$ \emph{weakly induced} by $S$ is defined as
$\WeaklyInduced{S}{G}=(S\cup N_G(S), E_G\cap (S\times (S\cup N_G(S))))$. 
For every graph $G$ and every $S\subseteq V_G$, $\subgraph{S}{G}$ 
is a subset of $\WeaklyInduced{S}{G}$, as $\WeaklyInduced{S}{G}$
additionally contains the neighbourhood of $S$ and every edge between elements
of $S$ and their neighbours. Also notice that $\WeaklyInduced{S}{G}$
does not contain the edges linking pairs of elements of $N_G(S)$.

An \emph{isomorphism} between two graphs $G=(V, E)$ and $G'=(V', E')$
is a bijective function $\varphi:V\to V'$ such that
$\forall_{v, w\in V}\ (v, w)\in E\iff (\varphi(v),\varphi(w))\in E'$.
Additionally, we denote by $\varphi(S)$ the restriction of $\varphi$
to a vertex subset $S\subseteq V$,
that is $\varphi(S)=\{\varphi(v)\mid v\in S\}$.

%---------------------------------
\subsection{Overview}
\label{subsec:overview}
%---------------------------------

\begin{figure*}[t]
\centering
\includegraphics[scale=1.1]{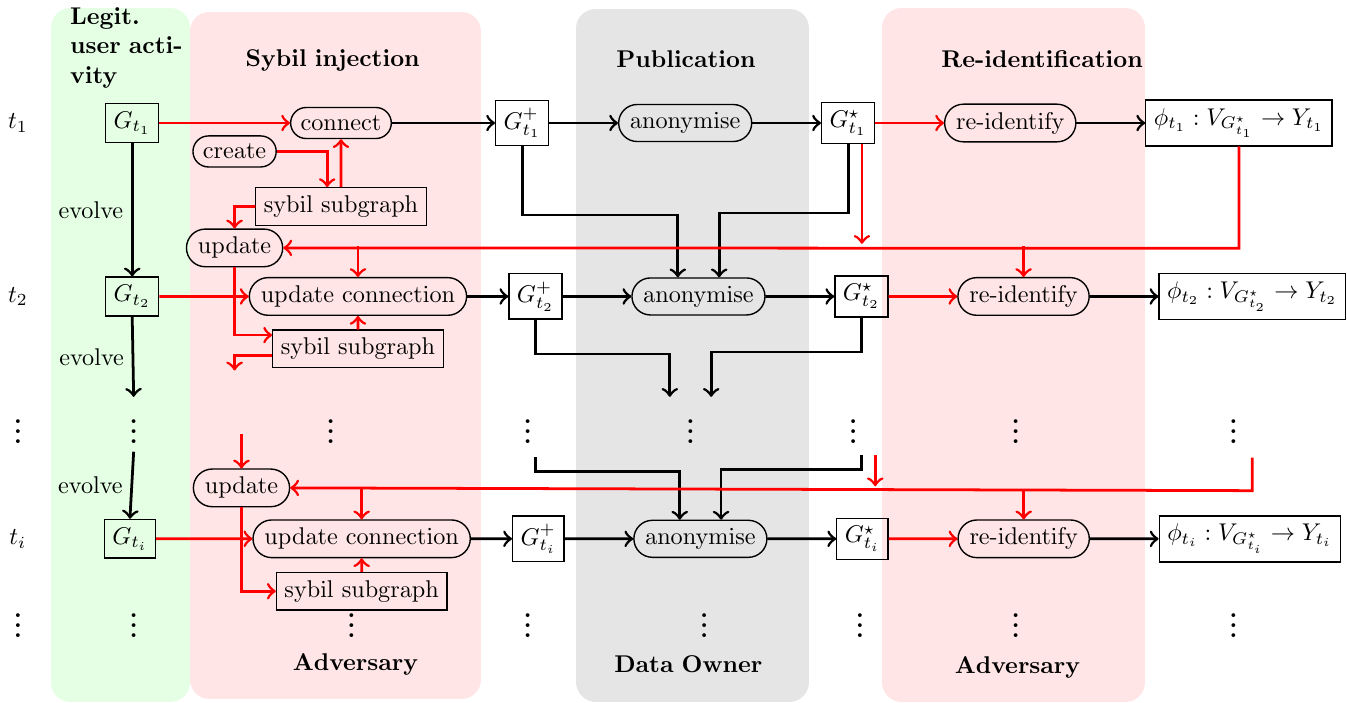}
\caption{Overview of the process of periodical graph publication
in the presence of active adversaries.}
\label{fig:overview}
\end{figure*}

Fig.~\ref{fig:overview} depicts the process of periodical graph publication
in the presence of an active adversary. We model this process as a game between
two players, the \emph{data owner} and the \emph{adversary}.
From a practical perspective, it is implausible for the data owner to release
a snapshot of the dynamic graph every time a small amount of changes occurs,
hence the periodical nature of the publication process. The data owner
selects a set of time-stamps $T=\{t_1, t_2, \ldots, t_i, \ldots\}$,
$t_1<t_2<\ldots<t_i<\ldots$, and incrementally publishes the sequence
$$\mathcal{G}^{\star}=(\Grat{t_1}, \Grat{t_2}, \ldots, \Grat{t_i}, \ldots)$$
of sanitised snapshots of the underlying dynamic social graph.
The adversary's goal is to re-identify, in a subset $T'
%=\{t_{i_1}, t_{i_2}, \ldots, t_{i_p}, \ldots\}
\subseteq T$
of the releases, a (possibly evolving) set of legitimate users
referred to as the \emph{victims}.
To achieve this goal, the active adversary injects an (also evolving) set
of fake accounts, commonly called \emph{sybils}, in the graph.
The sybil accounts create connections among themselves, and with the victims.
The connection patterns between each victim and some of the sybil nodes
is used as a unique \emph{fingerprint} for that victim.
The likely unique patterns built by the adversary
with the aid of the sybil nodes will enable her to effectively and efficiently
re-identify the victims in the sanitised snapshots.
At every re-identification attempt, the adversary first re-identifies
the set of sybil nodes, and then uses the fingerprints to re-identify the victims.

The data owner and the adversary have different partial views of the dynamic
social graph. On the one hand, the data owner knows the entire set of users,
both legitimate users and sybil accounts, but she cannot distinguish
them. The data owner also knows all relations. On the other hand, the adversary 
knows the identity of her victims and the structure of the subgraph weakly 
induced by the set of sybil nodes, but she does not know the structure 
of the rest of the network. 
In this paper we conduct the analysis
from the perspective of an external observer who can view all of the information.
For the sake of simplicity in our analysis, we will differentiate the sequence
$$\mathcal{G}^+=(\Gsat{t_1}, \Gsat{t_2}, \ldots, \Gsat{t_i}, \ldots),$$
which represents the view of the network according to the data owner,
which is the real network, i.e. the one containing the nodes representing
all users, both legitimate and malicious, from the sequence
$$\mathcal{G}=(\Gat{t_1}, \Gat{t_2}, \ldots, \Gat{t_i}, \ldots),$$
which represents the view of the unattacked network,
that is the view of the dynamic subgraph induced in $\mathcal{G}^+$
by the nodes representing legitimate users.

In the original formulation of active attacks, a single snapshot of the graph
is released, so all actions executed by the sybil nodes are assumed to occur
before the publication. This is not the case in the scenario
of a periodically released dynamic social graph. Here, the adversary
has the opportunity to schedule actions in such a way that the subgraph
induced by the sybil nodes evolves, as well as the set of fingerprints. In turn,
that allows her to use temporal patterns in addition to structural patterns
for re-identification. Additionally, the adversary can target different sets
of victims along the publication process and adapt the induced tempo-structural
patters to the evolution of the graph and the additional knowledge
acquired in each re-identification attempt.
In the new scenario, the actions performed by the adversary and the data owner
alternate as follows before, during and after each time-stamp $t_i\in T$. 

\vspace{2mm}\noindent
\textbf{Before $t_i$:} The adversary may remain inactive, or she can modify
the set of sybil nodes, as well the set of sybil-to-sybil
and sybil-to-victim edges. The result of these actions is the graph
$\Gsat{t_i}=(V_{t_i}\cup S_{t_i}, E_{t_i}\cup E^+_{t_i})$,  
where $V_{t_i}$ is the current set of legitimate users, $S_{t_i}$
is the current set of sybil nodes, $Y_{t_i}\subseteq V_{t_i}$ is the current
set of victims, $E_{t_i}=E_{G_{t_i}}\subseteq V_{t_i}\times V_{t_i}$
is the set of connections between legitimate users,
and $E^+_{t_i}\subseteq (S_{t_i}\times S_{t_i}) \cup (S_{t_i}\times Y_{t_i})$
is the set of connections created by the sybil accounts.
The subgraph $\WeaklyInduced{S_{t_i}}{\Gsat{t_i}}$, weakly induced in $\Gsat{t_i}$
by the set of sybil nodes, is the \emph{sybil subgraph}.
We refer to the set of modifications of the sybil subgraph executed
before the adversary has conducted any re-identification attempt
as \emph{sybil subgraph creation}. If the adversary has conducted
a re-identification attempt on earlier snapshots,
we refer to the modifications of the sybil subgraph
as \emph{sybil subgraph update}.

\vspace{2mm}\noindent
\textbf{During $t_i$}: The data owner applies an anonymisation method 
to $\Gsat{t_i}$
to obtain the sanitised version $\Grat{t_i}$, which is then released.
The anonymisation must preserve the consistency of the pseudonyms.
That is, every user must be labelled with the same pseudonym throughout
the sequence of snapshots where it appears. Consistent annotation
is of paramount importance for a number of analysis
tasks such as community evolution analysis~\cite{DTSB19},
link prediction~\cite{LZ11}, link persistence analysis~\cite{PK19},
among others, that require to track users along the sequence of releases.
The data owner anonymises every snapshot exactly once.

\vspace{2mm}\noindent
\textbf{After $t_i$}: The adversary adds $\Grat{t_i}$ to her knowledge.
At this point, she can remain inactive, or she can execute a re-identification
attempt on $\Grat{t_i}$. The result of a re-identification attempt
is a mapping $\phi_{t_i}: V_{\Grat{t_i}}\rightarrow Y_{t_i}$ determining
the pseudonyms assigned to the victims by the anonymisation method.
Here, the adversary can additionally modify the results of a previous
re-identification attempt conducted on some of the preceding releases.

%---------------------------------------------------
\subsection{Components of the Process}
\label{ssec:game-components}
%---------------------------------------------------

To discuss in detail the different actions that the data owner or the adversary
execute, we follow the categorisation given in above in terms of the time
where each action may occur for every time-stamp $t_i$. We first discuss
sybil subgraph creation and update, which occur before $t_i$,
then graph publication, which occurs at $t_i$, and finally re-identification,
which occurs after $t_i$.

\subsubsection{Sybil subgraph creation and update}

As we mentioned above, sybil subgraph creation is executed before the adversary
has attempted re-identification for the first time; whereas sybil subgraph
update is executed in the remaining time-steps.

\vspace{2mm}\noindent
{\bf Sybil subgraph creation:}
In the dynamic scenario, the adversary can build the initial sybil subgraph 
along several releases.
This allows the creation of tempo-structural patterns, incorporating
information about the first snapshot where each sybil node appears,
to facilitate the sybil subgraph retrieval stage during re-identification.
As in all active attacks, the patterns created must ensure that,
with high probability, $\Gsb{t_i}$ is unique. 
We denote by $F_{t_i}(y)$ the fingerprint of a victim $y\in Y_{t_i}$ in terms
of $S_{t_i}$. Throughout this paper we consider that $F_{t_i}(y)$
is uniquely determined by the neighbourhood of $y$ in $S_{t_i}$, that is
$F_{t_i}(y)=S_{t_i}\cap N_{\Gsat{y}}
%=\{v\in S_{t_i}\;|\;(v,y)\in E^+_{t_i}\}
$. 
We denote by $\mathcal{F}_{t_i}$ the set of fingerprints of all victims
in~$\Gsat{t_i}$.

\vspace{2mm}\noindent
{\bf Sybil subgraph update:}
In this step, the adversary can modify the set of sybil nodes, by adding new
sybil nodes or replacing existing ones. The latter action helps to make
every individual sybil node's behaviour appear more normal,
which in turn makes it more likely to succeed in establishing
the necessary relations and less likely to be detected
by sybil defences. The adversary can also modify the inter-sybil connections
and the fingerprints. Since sybil subgraph update is executed after
at least one re-identification
attempt has been conducted, the adversary can use information from this attempt,
such as the level of uncertainty in the re-identification,
to decide the changes to introduce in the sybil subgraph.
Finally, if the number of fingerprints that can be constructed using
the new set of sybil nodes is larger than the previous number of targeted victims,
that is $2^{|S_{t_i}|}-1>|Y_{t_{i-1}}|$, the adversary can additionally target
new victims, either new users that joined the network in the last inter-release
interval, or previously enrolled users that had not been targeted so far.
In the latter case, even if these victims had not been targeted before,
the consistency of the labelling in the sequence of sanitised snapshots
entails that a re-identification in the $t_i$-th snapshot can be traced back
to the previous ones.

\subsubsection{Graph publication}

As we discussed above, at time step $t_i$, the data owner anonymises $\Gsat{t_i}$
and publishes the sanitised version $\Grat{t_i}$. From a general point of view,
we treat the anonymisation as a two-step process. The first step
is \emph{pseudonymisation}, which consists in building the isomorphism
$\varphi_{t_i}:V_{\Gsat{t_i}}\rightarrow V'^{\star}_{t_i}$,
with $V'^{\star}_{t_i}\cap V_{\Gsat{t_i}}=\emptyset$,
that replaces every real identity in $\Gsat{t_i}$ for a pseudonym.
The pseudonymised graph is denoted as $\varphi_{t_i}\Gsat{t_i}$.
If $i=1$, all pseudonyms are freshly generated. In the remaining cases,
the pseudonyms for previously existing vertices are kept,
and fresh pseudonyms are assigned to new vertices.

Releasing the pseudonymised graph has been proven to be insufficient
for preventing re-identification \cite{NS09,LT08,TY2016,BDK11,MRT19}.
Thus, the second step of the anonymisation process consists in applying
a \emph{perturbation} method $\Phi_{t_i}: \varphi_{t_i}\Gsat{t_i}\rightarrow
(V^{\star}_{t_i}, V^{\star}_{t_i} \times V^{\star}_{t_i})$
to the pseudonymised graph. Perturbation consists in editing the vertex
and/or edge sets of the pseudonymised graph in such a way that the resulting graph
satisfies some privacy guarantee against re-identification.
For the case of active adversaries, the relevant perturbation methods
are the ones based on random vertex/edge flipping and those
based on the notions of $(k,\Gamma_\ell)$-(adjacency) anonymity
\cite{TY2016,MRT18,MRT18b,MRT19}.
Finally, the data owner releases the graph $\Grat{t_i}$ obtained
as the result of applying pseudonymisation on $\Gsat{t_i}$ and perturbation
on $\varphi_{t_i}\Gsat{t_i}$,
that is $\Grat{t_i}=\Phi_{t_i}(\varphi_{t_i}\Gsat{t_i})$.

\subsubsection{Re-identification}

In the new scenario, re-identification is dynamic, as it occurs along several
time-steps, leveraging the increase of the adversary knowledge after the release
of every new snapshot. Considering the applicable techniques, we differentiate
the first re-identification attempt, which can be executed
immediately after the publication of $\Grat{t_i}$, from the remaining attempts,
which we refer to as refinements, and can be executed after the release
of every other $\Grat{t_j}$, $j>i$.

\vspace{2mm}\noindent
{\bf First re-identification attempt:}
The first re-identification attempt is composed of two steps:
sybil subgraph retrieval and fingerprint matching.
From a general point of view, the procedure consists in the following steps:
\begin{enumerate}
\item Sybil subgraph retrieval:
\begin{enumerate}
\item Find in $\Grat{t_i}$ a set $\mathcal{X}_{t_i}=\{X_1, X_2, \ldots, X_p\}$,
$X_j\subseteq V_{\Grat{t_i}}$, of candidate sybil sets.
For every $X\in \mathcal{X}_{t_i}$, the graph $\WeaklyInduced{X}{\Grat{t_i}}$
is a candidate sybil subgraph. Every specific attack defines
the conditions under which a candidate is added to~$\mathcal{X}_{t_i}$.
\item Filter out elements of $\mathcal{X}_{t_i}$
that fail to satisfy temporal consistency constraints
with respect to $\Gsb{t_1}$, $\Gsb{t_2}$, \ldots, $\Gsb{t_{i-1}}$.
The specific constraints to enforce depend on the instantiation
of the attack strategy.
\item If $\mathcal{X}_{t_i}=\emptyset$, the attack fails.
Otherwise, proceed to fingerprint matching (step 2).
\end{enumerate}
\item Fingerprint matching:
\begin{enumerate}
\item Select one element $X\in \mathcal{X}_{t_i}$. As in the previous steps,
every specific attack defines how the selection is~made.
\item Using $X$ and $\mathcal{F}_{t_i}$, find a set of candidate
mappings $\mathcal{Y}_{X}=\left\{\phi_1, \phi_2, \ldots, \phi_q \right\}$,
where every $\phi_j$ ($1\le j \le q$) has the form
$\phi_j:V_{\Grat{t_i}}\setminus S_{t_i}\rightarrow Y_{t_i}$.
Every element of $\mathcal{Y}_{X_{t_i}}$ represents a possible re-identification
of the victims in $\Grat{t_i}$.
\item Filter out elements of $\mathcal{Y}_{X_{t_i}}$
that fail to satisfy (attack-specific) temporal consistency constraints
with respect to
$\mathcal{F}_{t_1}, \mathcal{F}_{t_2}, \ldots, \mathcal{F}_{t_{i-1}}$.
\item If $\mathcal{Y}_{\mathcal{X}'_{t_i}}=\emptyset$, the attack fails.
Otherwise, select one element of $\mathcal{Y}_{\mathcal{X}'_{t_i}}$
and give it as the result of the re-identification. As in the previous steps,
every specific attack defines how the selection is made.
\end{enumerate}
\end{enumerate}

In an actual instantiation of the attack, steps 1.a-c,
as well as steps 2.a-d, are not necessarily executed in that order,
nor independently. As we will show in Sects.~\ref{sec:attack}
and~\ref{sec:experiments}, combining temporal consistency constraints
with structural similarity allows for higher effectiveness
and considerable speed-ups in several steps.

\vspace{2mm}\noindent
{\bf Re-identification refinement:}
As we discussed above, the first re-identification attempt on $\Grat{t_i}$
can be executed immediately after the snapshot is published. Then, after
the publication of $\Grat{t_j}$, $j>i$, the re-identification refinement step
allows to improve the adversary's certainty on the previous re-identification,
by executing the following actions:
\begin{enumerate}
\item Filter out elements of $\mathcal{X}_{t_i}$ that fail to satisfy
additional temporal consistency constraints with respect to $\Gsb{t_i}$
and $\Gsb{t_j}$.
\item Repeat the fingerprint matching step.
\end{enumerate}

Note that, in a specific attack, the adversary may choose to run
the re-identification on $\Grat{t_i}$ only once, waiting for the release
of several $\Grat{t_{j_1}}$, $\Grat{t_{j_2}}$, \ldots, $\Grat{t_{j_r}}$,
and combining all temporal consistency checks of the first attempt
and the refinements in a single execution of step 1.b described above.
However, we keep the main re-identification attempt separable from the refinements
considering that in a real-life attack the actual time elapsed
between $\Grat{t_i}$ and $\Grat{t_j}$ can be considerably large,
e.g. several months.

%===================================================
\section{A Novel Dynamic Active Attack}
\label{sec:attack}
%===================================================

In this section we present what is, to the best of our knowledge,
the first active attack on periodically released dynamic social graphs.
The novelty of our attack lies in its ability to exploit the dynamic nature
of the social graph being periodically published, and the fact
that the publication process occurs incrementally. Our attack benefits
from temporal information in two fundamental ways. Firstly, we define
a number of temporal consistency constraints, and use them in all stages
of the re-identification process. In sybil subgraph retrieval,
consistency constraints allow us to obtain considerably small sets
of plausible candidates, which increases the likelihood
of the attacker selecting the correct one. A similar situation occurs
in fingerprint matching. Moreover, the incremental publication process
allows the adversary to refine previous re-identifications by applying
new consistency checks based on later releases. In all cases, temporal
consistency constraints additionally make the re-identification significantly
fast, especially when compared with comparably noise-resilient methods
reported in the literature for the static publication scenario. The second manner
in which our attack benefits from the dynamicity of the publication process
is by adapting the set of fingerprints in the interval between consecutive
re-identification attempts in such a way that the level of uncertainty
in the previous re-identification is reduced.

In the remainder of this section, we will describe out new attack in detail.
We will first introduce the notions of temporal consistency.
Then, we will describe the manner
in which temporal consistency is exploited for dynamic re-identification.
Finally, we will describe how tempo-structural patterns
are created and \mbox{maintained}.

\subsection{Temporal consistency constraints}
\label{ssec:temp-consistency}

As we discussed in Sect.~\ref{subsec:overview}, the data owner must assign
the same pseudonym to each user throughout the subsequence of snapshots
where it appears, to allow for analysis tasks such as community evolution
analysis~\cite{DTSB19}, link prediction~\cite{LZ11}, link persistence
analysis~\cite{PK19}, etc. Since the data owner cannot distinguish
between legitimate users (including victims) and sybil accounts,
she will assign time-persistent pseudonyms to all of them.
Additionally, since the adversary receives all sanitised snapshots,
she can determine when a pseudonym was used for the first time,
whether it is still in use, and in case it is not, when it was used
for the last time.

In our attack, the adversary exploits this information in all stages
of the re-identification process. For example, consider the following situation.
The set of sybil nodes at time-step $t_6$ is $S_{t_6}=\{s_1, s_2, s_3, s_4\}$.
The adversary inserted $s_1$ and $s_2$ in the interval
preceding the publication of $\Grat{t_2}$. Additionally, she inserted $s_3$
before the publication of $\Grat{t_3}$ and $s_4$ before the publication
of $\Grat{t_5}$. After the release of $\Grat{t_6}$, during the sybil subgraph
retrieval phase of the first re-identification attempt, the adversary needs
to determine whether a set $X\subseteq V_{\Grat{t_6}}$,
say $X=\{v_1,v_2,v_3,v_4\}$, is a valid candidate.
Looking at the first snapshot where each of these pseudonyms was used,
the adversary observes that $v_1$ and $v_3$ were first used in $\Grat{t_2}$,
so they are feasible matches for $s_1$ and $s_2$, in some order.
Likewise, $v_2$ was first used in $\Grat{t_5}$, so it is a feasible match
for $s_4$. However, she observes that $v_4$ was first used in $\Grat{t_4}$,
unlike any element of $S_{t_6}$. From this observation, the adversary knows
that $X$ is not a valid candidate, regardless of how structurally similar
$\WeaklyInduced{X}{\Grat{t_6}}$ and $\WeaklyInduced{S_{t_6}}{\Gsat{t_6}}$ are.

The previous example illustrates how temporal consistency constraints
are used for discarding candidate sybil sets. We now formalise the different
types of constraints used in our attack. To that end, we introduce
some new notation. The function
$\alpha^+:\cup_{t_i\in T} V_{\Gsat{t_i}}\rightarrow T$
yields, for every vertex $v\in\cup_{t_i\in T} V_{\Gsat{t_i}}$, the order
of the first snapshot where $v$ exists, that is
$$\alpha^+(v)=\min \{\{t_i\in T\;|\;v\in V_{\Gsat{t_i}}\}\}.$$
Analogously, the function
$\alpha^\star:\cup_{t_i\in T} V_{\Grat{t_i}}\rightarrow T$
yields the order of the first snapshot where each pseudonym is used,
that is $$\alpha^\star(x)=t_i\iff
\exists_{v\in V_{\Gsat{t_i}}}\ \alpha^+(v)=t_i\wedge \varphi_{t_i}(v)=x.$$

Clearly, the adversary knows the values of the function~$\alpha^\star$
for all pseudonyms used by the data owner. Additionally, she knows
the values of $\alpha^+$ for all of her sybil nodes. Thus, the previous functions
allow us to define the notion of \emph{first-use-as-sybil consistency},
which is used by the sybil subgraph retrieval method.

\begin{definition}\label{def-first-use-as-sybil}
Let $X\subseteq V_{\Grat{t_i}}$ be a set of pseudonyms such that $|X|=|S_{t_i}|$
and let $\phi:S_{t_i}\rightarrow V_{\Grat{t_i}}$ be a mapping from the set
of real sybil nodes to the elements of $X$. We say that $X$ and $S_{t_i}$
satisfy \emph{first-use-as-sybil consistency} according to $\phi$,
denoted as $X \simeq_{\phi} S_{t_i}$,
if and only if $\forall_{s\in S_{t_i}}\ \alpha^+(s)=\alpha^\star(\phi(s))$.
\end{definition}

Note that first-use-as-sybil consistency depends on the order
in which the elements of the candidate set are mapped to the real sybil nodes,
which is a requirement of the sybil subgraph retrieval method.

We define an analogous notion of first use consistency for victims.
In this case, the adversary may or may not know the value of $\alpha^+$.
In our attack, we assume that she does not, and introduce an additional function
to represent the temporal information the adversary must necessarily have
about victims. The function $\beta^+:\cup_{t_i\in T} Y_{t_i} \rightarrow T$
yields, for every $v\in \cup_{t_i\in T} Y_{t_i}$, the order of the snapshot
where $v$ was targeted for the first time, that~is
$$\beta^+(v)=\min \{\{t_i\in T\;|\;v\in Y_{t_i}\}\}.$$

The new function allows us to define the notion
of \emph{first-time-targetted consistency}, which is used
in the fingerprint matching method.

\begin{definition}\label{def-first-time-targetted}
Let $v\in V_{\Grat{t_i}}$ be a victim candidate and let $y\in Y_{t_i}$
be a real victim. We say that $v$ and $y$
satisfy \emph{first-time-targetted consistency}, denoted as $v\simeq y$,
if and only if $\alpha^\star(v)\le\beta^+(y)$.
\end{definition}

This temporal consistency notion encodes the rationale that the adversary
can ignore during fingerprint matching those pseudonyms that the data owner
used for the first time after the corresponding victim had been targeted.

Finally, we define the notion of \emph{sybil-removal-count consistency},
which is used by the re-identification refinement method to encode the rationale
that a sybil set candidate $X$, for which no temporal inconsistencies were found
during the $t_i$-th snapshot, can be removed from $\mathcal{X}_{t_i}$
when the $t_{i+1}$-th snapshot is released, if the number of sybil nodes removed
by the adversary in the interval between these snapshots does not match the number
of elements of $X$ that cease to exist in $\Grat{t_{i+1}}$.

\begin{definition}\label{def-sybil-removal-count}
We say that a set of pseudonyms $X\subseteq V_{\Grat{t_i}}$ satisfies
\emph{sybil-removal-count consistency} with respect
to the pair $(S_{t_i},S_{t_{i+1}})$,
which we denote as $X \simeq (S_{t_i},S_{t_{i+1}})$,
if and only if $|X\setminus V_{\Grat{t_{i+1}}}|=|S_{t_i}\setminus S_{t_{i+1}}|$.
\end{definition}

Defs.~\ref{def-first-use-as-sybil} to~\ref{def-sybil-removal-count}
play an important role in the new dynamic re-identification methods
introduced as part of the new attack, as we will discuss in what follows.

%---------------------------------------------------
\subsection{Dynamic re-identification}

In what follows we describe the methods for sybil subgraph retrieval
and fingerprint matching, which are conducted during
the first re-identification attempt,
as well as the re-identification refinement method. In all cases,
we pose the emphasis on the manner in which they use the notions
of temporal consistency for maximising effectiveness and efficiency.

\subsubsection{Sybil subgraph retrieval}
\label{ssec:attack-sybil-subgraph-retrieval}

The sybil subgraph retrieval method is a breath-first search procedure,
which shares the philosophy of analogous methods devised for active attacks
on static graphs~\cite{BDK11,MRT19}, but differs from them in the use
of temporal consistency constraints for pruning the search space.
To establish the order in which the search space is traversed, our method
relies on the existence of an arbitrary (but fixed) total order $\prec$ among
the set of sybil nodes, which is enforced by the sybil subgraph creation
method and maintained by the sybil subgraph update method.

Let $s_1\prec s_2 \prec \ldots \prec s_{|S_{t_i}|}$ be the order established
on the elements of $S_{t_i}$. The search procedure first builds a set
of cardinality-$1$ partial candidates
$$\mathcal{X}_{t_i,1}=\{\{v_{j_1}\}\mid v_{j_1}\in V_{\Grat{t_i}}\}.$$
Then, it obtains the pruned set of candidates $\mathcal{X}'_{t_i,1}$
by removing from $\mathcal{X}_{t_i,1}$ all elements $\{v_{j_1}\}$
such that $\alpha^\star(v_{j_1})\ne\alpha^+(s_1)$,
or $\left|\delta_{\Grat{t_i}}(v_{j_1})-\delta_{\Gsat{t_i}}(s_1)\right|>\theta$.
The first condition verifies that the first-use-as-sybil consistency property
$\{v_{j_1}\}\simeq_{\phi} \{s_1\}$ holds, with $\phi=\{(s_1,v_{j_1})\}$.
The second condition is analogous to the one applied in the noise-tolerant
sybil subgraph retrieval method introduced in~\cite{MRT19}
as part of the so-called \emph{robust active attack}. It aims to exclude
from the search tree all candidates $X$ such that
$\Delta(\WeaklyInduced{X}{\Grat{t_i}},\WeaklyInduced{S_{t_i}}{\Gsat{t_i}})
>\theta$, where $\Delta$ is a structural dissimilarity function and $\theta$
is a tolerance threshold.

After pruning $\mathcal{X}_{t_i,1}$, the method builds the set
of cardinality-$2$ partial candidates
$$\mathcal{X}_{t_i,2}=\{\{v_{j_1},v_{j_2}\}\mid
\{v_{j_1}\}\in\mathcal{X}_{t_i,\ell},
v_{j_2}\in V_{\Grat{t_i}}\setminus\{v_{j_1}\}\}.$$
Similarly, $\mathcal{X}_{t_i,2}$ is pruned by removing all elements
$\{v_{j_1},v_{j_2}\}$ such that $\{v_{j_1}, v_{j_2}\}
\not\simeq_{\phi} \{s_1, s_2\}$, with $\phi=\{(s_1,v_{j_1}),(s_2,v_{j_2})\}$,
and $\Delta(\WeaklyInduced{\{v_{j_1},v_{j_2}\}}{\Grat{t_i}},
\WeaklyInduced{\{s_1, s_2\}}{\Gsat{t_i}})>\theta$.

In general, for $\ell\le|S_{t_i}|$, the method builds the set
of partial candidates
\begin{align*}
\mathcal{X}_{t_i,\ell}=\{\{v_{j_1},\ldots,v_{j_\ell}\}\mid &
\ \{v_{j_1}\ldots,v_{j_{\ell-1}}\}\in\mathcal{X}_{t_i,\ell-1},\\
&\ v_{j_\ell}\in V_{\Grat{t_i}}\setminus\{v_{j_1},\ldots,v_{j_{\ell-1}}\}\}
\end{align*}
and obtains the pruned candidate set $\mathcal{X}'_{t_i,\ell}$ by removing
from $\mathcal{X}_{t_i,\ell}$ all elements $\{v_{j_1},\ldots,v_{j_\ell}\}$
such that
$$\{v_{j_1},\ldots,v_{j_\ell}\}\not\simeq_{\phi} \{s_1,\ldots,s_\ell\},$$
with $\phi=\{(s_1,v_{j_1}),\ldots,(s_\ell,v_{j_\ell})\}$, and
$$\Delta(\WeaklyInduced{\{v_{j_1},\ldots,v_{j_\ell}\}}{\Grat{t_i}},
\WeaklyInduced{\{s_1,\ldots,s_\ell\}}{\Gsat{t_i}})>\theta.$$

In our attack, we use the structural dissimilarity measure defined
in~\cite{MRT19}, which makes
$$
\Delta(\WeaklyInduced{\{v_{j_1},\ldots,v_{j_\ell}\}}{\Grat{t_i}},
\WeaklyInduced{\{s_1,\ldots,s_\ell\}}{\Gsat{t_i}})=
|D|+\sum_{k=1}^{\ell}\left|\delta'_{\Gsat{t_i}}(s_k)
-\delta'_{\Grat{t_i}}(v_{j_k})\right|
$$
where
$\delta'_{\Gsat{t_i}}(s_k)$ is is the number of neighbours of $s_k$
in $\Gsat{t_i}$ that are not in $\{s_1,\ldots,s_\ell\}$,
$\delta'_{\Grat{t_i}}(v_{j_k})$ is is the number of neighbours of $v_{j_k}$
in $\Grat{t_i}$ that are not in $\{v_{j_1},\ldots,v_{j_\ell}\}$, and
$$|D|=|\{(j_k,j_{k'})\mid |E_{\Grat{t_i}}\cap\{(v_{j_k},v_{j_{k'}})\}| 
+|E_{\Gsat{t_i}}\cap\{(s_k,s_{k'})\}|=1\}|$$
is an efficiently computable estimation
of the edge-edit distance
between $\subgraph{\{v_{j_1},\ldots,v_{j_\ell}\}}{\Grat{t_i}}$
and $\subgraph{\{s_1,\ldots,s_\ell\}}{\Gsat{t_i}}$.

Finally, the sybil subgraph retrieval method gives as output
the pruned set of cardinality-$|S_{t_i}|$ candidates, that~is
$$\mathcal{X}_{t_i}=\mathcal{X}'_{t_i,|S_{t_i}|}.$$
In other words, our method gives as output the set of temporally
consistent subsets whose weakly induced subgraphs in $\Grat{t_i}$
are structurally similar, within a tolerance threshold $\theta$,
to that of the original set of sybil nodes in~$\Gsat{t_i}$.

\subsubsection{Fingerprint matching}
\label{ssec:attack-fp-matching}

After $\mathcal{X}_{t_i}$ is obtained, a candidate
%$\WeaklyInduced{X}{\Grat{t_i}}$,
$X=\{v_{j_1}, v_{j_2},$ $\ldots v_{j_{|S_{t_i}|}}\}$ is randomly selected
from $\mathcal{X}_{t_i}$, with probability $\frac{1}{|\mathcal{X}_{t_i}|}$,
for conducting the fingerprint matching step.
Let $v_{j_1} \prec v_{j_2} \prec \ldots \prec v_{j_{|S_{t_i}|}}$ be the order
established on the elements of~$X$ by the sybil subgraph retrieval method.

Our fingerprint matching method is a depth-first search procedure,
which gives as output
a set $\mathcal{Y}_{X}=\{\phi_1, \phi_2, \ldots, \phi_q\}$,
where every $\phi\in\mathcal{Y}_X$ has the form
$\phi: Y_{t_i}\rightarrow N_{\Grat{t_i}}(X)$.
Every element of $\mathcal{Y}_{X}$ maximises the pairwise similarities
between the original fingerprints of the victims and the fingerprints,
with respect to $X$, of the corresponding pseudonymised vertices.

The method first finds all equally best matches
between the (real) fingerprint $F_j$ of a victim $y_j\in Y_{t_i}$ and that
of a temporally consistent vertex $u\in N_{\Grat{t_i}}(X)$ with respect to $X$,
that~is $F^\star_u=N_{\Grat{t_i}}(u)\cap X$.
Then, for every such match, it recursively applies the search procedure
to match the remaining real victims to other temporally consistent
candidate victims. For every victim $y_j$ and every candidate match $u$,
the similarity function $\similarity(F^\star_u,F_j)$ integrates the verification
of the temporal consistency and the structural fingerprint,
and is computed~as
\begin{displaymath}
\similarity(F^\star_u,F_j)=\left\{
\begin{array}{ll}
\similarity_c(F^\star_u,F_j)&\text{if }u\simeq y_j\\
&\text{and }\similarity_c(F^\star_u,F_j)\ge\eta\\
0&\text{otherwise.}
\end{array}
\right.
\end{displaymath}
where $\eta$ is a tolerance threshold allowing to ignore
insufficiently similar matches and the function
$\similarity_c(F^\star_u,F_j)$ is defined as
$$\similarity_c(F^\star_u,F_j)=\sum_{k=1}^{|S_{t_i}|}
\mu_{k}(F^\star_u,F_j)$$
with
\begin{displaymath}
\mu_{k}(F^\star_u,F_j)=\left\{
\begin{array}{ll}
1&\text{ if }v_{j_k}\in F^\star_u\text{ and }s_k\in F_j \\
0&\text{ otherwise}.
\end{array}
\right.
\end{displaymath}

Our method is similar to the one introduced in~\cite{MRT19}
in the fact that it discards matchings whose structural similarity
is insufficiently high. Moreover, temporal consistency constraints
allow our method to considerably reduce
the number of final candidate \mbox{mappings}.

%---------------------------------------------------
\subsubsection{Re-identification refinement}
\label{ssec:attack-re-ident-refinement}
%---------------------------------------------------

After the $t_{i+1}$-th snapshot is released, the adversary obtains additional
information that can improve the re-identification at the $t_i$-th snapshot.
In specific, the adversary learns
the set $V_{\Grat{t_i}}\setminus V_{\Grat{t_{i+1}}}$
of pseudonyms corresponding to users that ceased to be members
of the social network in the interval between the $t_i$-th
and the $t_{i+1}$-th snapshots. If the adversary removed some sybil nodes
$s_{j_1}$, $s_{j_2}$, \ldots, $s_{j_r}$ in this interval, then she knows
that $\{\varphi_{t_i}(s_{j_1}), \varphi_{t_i}(s_{j_2}), \ldots,
\varphi_{t_i}(s_{j_r})\}\subseteq V_{\Grat{t_i}}\setminus V_{\Grat{t_{i+1}}}$.
This information allows the adversary to refine the set $\mathcal{X}_{t_i}$
obtained in the first re-identification attempt on $\Grat{t_i}$.
Certainly, a candidate $X$
such that $|X\setminus V_{\Grat{t_{i+1}}}|\ne|S_{t_i} \setminus S_{t_{i+1}}|$
is not a valid match for $S_{t_i}$.

Thus, after the publication of $\Grat{t_{i+1}}$, the adversary refines
$\mathcal{X}_{t_i}$ by removing
the candidates that violate the sybil-removal-count consistency notion, that is
$$\mathcal{X}'_{t_i}=\mathcal{X}_{t_i}
\setminus\left\{X\mid X\not\simeq (S_{t_i},S_{t_{i+1}})\right\},$$
and re-runs the fingerprint matching step with $\mathcal{X}'_{t_i}$.
%if~necessary.

\subsection{Sybil Subgraph Creation and Update}

Here we describe how the necessary tempo-structural patterns
for dynamic re-identification are created and maintained.
Both the sybil subgraph creation and the sybil subgraph update stages contribute
to this task. Additionally, the sybil subgraph update addresses
other aspects of the dynamic behaviour of the new attack, e.g. targeting
new victims.

\subsubsection{Sybil subgraph creation}
\label{ssec:attack-initial-ssg}

In the dynamic attack, the initial sybil graph is not necessarily created
before the first snapshot is released. Let $\Grat{t_i}$ be the first snapshot
where the adversary conducts a re-identification attempt. Then, the sybil
sugbraph creation is executed during the entire time window preceding $t_i$.
The adversary initially inserts a small number of sybil nodes, no more than
$\left\lfloor\log_2 \left(|V_{\Gsat{t_i}}|\right)\right\rfloor$.
This makes the sybil subgraph very unlikely to be detected
by sybil defences~\cite{Yu2006,Yu2008,BDK11,MRT18b,MRT19}, while allowing
to create unique fingerprints for a reasonably large number of potential
initial victims. Spreading sybil injection over several snapshots
helps create temporal patterns that reduce the search space
during sybil subgraph retrieval.

As sybil nodes are inserted, they are connected to other sybil nodes
and to some of the victims. Inter-sybil edges are created
in a manner that has been shown in~\cite{BDK11} to make
the sybil subgraph unique with high probability, which helps in accelerating
the breath-first search procedure in the basis of sybil subgraph retrieval.
First, an arbitrary (but fixed) order is established among the sybil nodes.
In our case, we simply take the order in which the sybils are created.
Let $s_1 \prec s_2 \prec \ldots \prec s_{|S_{t_i}|}$ represent
the order established among the sybils. Then, the edges $(s_1,s_2)$, $(s_2,s_3)$,
\ldots, $(s_{|S_{t_i}|-1}, s_{|S_{t_i}|})$ are added to force the existence
of the path $s_1 s_2\ldots s_{|S_{t_i}|}$.
Additionally, every other edge $(s_j, s_k)$, $|j-k|\ge 2$, is added
with probability~$0.5$. 
The initial fingerprints of the elements of $Y_{t_i}$ are
randomly generated by connecting
each victim to each sybil node with probability $0.5$,
checking that all fingerprints are unique.

\subsubsection{Sybil subgraph update}
\label{ssec:attack-ssg-update}

Let $\Grat{t_{i-1}}$ and $\Grat{t_i}$ be two consecutive releases occurring
after the first snapshot where the adversary conducted a re-identification attempt
($\Grat{t_{i-1}}$ itself may have been this snapshot).
In the interval between $\Grat{t_{i-1}}$ and $\Grat{t_i}$, the adversary updates
the sybil subgraph by adding and/or removing sybil nodes and inter-sybil edges,
updating the fingerprints of (a subset of) the victims,
and possibly targeting new victims. The changes made in the sybil subgraph
aim to improve the sybil subgraph retrieval and fingerprint matching steps
in future re-identification attempts. We describe each of these modifications
in detail in what follows.

\subsubsection*{Adding and replacing sybil nodes}
\label{sssec:attack-sybil-set-update}

In our attack, the adversary is conservative regarding the number of sybil nodes,
balancing the capacity to target more victims with the need to keep
the likelihood of being detected by sybil defences sufficiently low.
Thus, the number of sybil nodes is increased as the number of nodes in the graph
grows, but keeping
$|S_{t_i}|\le \left\lfloor\log_2 \left(|V_{\Grat{t_{i-1}}}|\right)\right\rfloor$.
If the graph growth rate between releases is small,
this strategy translates into not increasing the number of sybils
during many consecutive releases. However, this does not mean
that new sybil nodes are not created, since the attack additionally
selects a (small) random number of existing sybil nodes and replaces
them for fresh sybil nodes. The purpose of these replacements is twofold.
First, they allow to keep the activity level of each individual sybil node
sufficiently low, and thus make it less distinguishable from legitimate nodes.
Secondly, the frequent modification of the set of sybils helps reduce
the search space for sybil subgraph retrieval, by increasing the number
of potential candidates violating the first-use-as-sybil consistency constraint,
and enables the re-identification refinement process to detect more sybil
set candidates violating the sybil-removal-count consistency constraint.

We now discuss the changes that the adversary does in the set of inter-sybil
edges to handle sybil node addition and replacement.
Let $S_{t_{i-1}}=\{s_1, s_2, \ldots, s_{|S_{t_{i-1}}|}\}$ be the set of sybil
nodes present in $\Gsat{t_{i-1}}$,
and let $s_1 \prec s_2 \prec \ldots \prec s_{|S_{t_{i-1}}|}$
be the order established among them. We first consider the case of sybil node
addition, where no existing sybil node is replaced.
Let $S'=\{s'_1, s'_2, \ldots, s_{q}\}$ be the set of new sybil nodes
that will be added to $\Gsat{t_i}$,
and let $s'_1 \prec s'_2 \prec \ldots \prec s_{q}$
be the order established on them. A structure analogous to that of the previous
sybil subgraph is enforced by adding to $\Gsat{t_i}$
the edges $(s_{|S_{t_{i-1}}|}, s'_1)$, $(s'_1, s'_2)$, \ldots, $(s'_{q-1}, s'_q)$,
which results in extending the path $s_1 s_2 \ldots s_{|S_{t_{i-1}}|}$
into $s_1 s_2 \ldots s_{|S_{t_{i-1}}|} s'_1 s'_2 \ldots s_{q}$.
Additionally, the adversary adds to $\Gsat{t_i}$ every node $(x,y)$,
$x\in S'$, $y\in (S_{t_{i-1}}\cup S')\setminus N_{\Gsat{t_i}}(x)$,
with probability $0.5$. 

Now, we describe the modifications made by the adversary for replacing
a sybil node $s_j\in S_{t_{i-1}}$ for a new sybil node $s$ ($s\notin S'$).
In this case, the adversary adds to $\Gsat{t_i}$ the edges $(s_{j-1},s)$
and $(s,s_{j+1})$, where $s_{j-1}$ and $s_{j+1}$ are the sybil nodes
immediately preceding and succeeding $s_j$ according to $\prec$.
The order $\prec$ is updated accordingly to make $s_1 \prec s_2 \prec \ldots
\prec s_{j-1} \prec s \prec s_{j+1} \prec \ldots \prec s_{|S_{i-1}|}$.
These modifications ensure that the path
$s_1s_2\ldots s_{|S_{t_{i-1}}|}$ guaranteed to exist in $\Gsat{t_{i-1}}$
is replaced in $\Gsat{t_i}$
for $s_1s_2\ldots s_{j-1}ss_{j+1}\ldots s_{|S_{i-1}|}$.
Additionally the new sybil node $s$ is connected to every other sybil node
with probability $0.5$.
In our attack, every sybil node removal is part of a replacement,
so the number of sybil nodes never decreases.

\subsubsection*{Updating fingerprints of existing victims}
\label{sssec:attack-fp-update}

After replacing a sybil node $s\in S_{t_{i-1}}$ for a new sybil node
$s'\in S_{t_i}\setminus S_{t_{i-1}}$, the adversary adds to $\Gsat{t_i}$
the edge $(s',y)$ for every $y\in Y_{t_{i-1}}\cap N_{\Gsat{t_{i-1}}}(s)$,
to guarantee that the replacement of $s$ for $s'$ does not render
any pair of fingerprints identical in $\Gsat{t_i}$.

Additionally, if new sybil nodes were added, the fingerprints
of all previously targeted victims in $Y_{t_{i-1}}$ are modified
by creating edges linking them to a subset of the new sybil nodes.
For each new sybil node $s\in S_{t_i}\setminus S_{t_{i-1}}$
and every victim $y\in Y_{t_{i-1}}$, the edge $(s,y)$ is added
with probability $0.5$.

Finally, if the adversary has conducted a re-identification attempt
on $\Grat{t_{i-1}}$, she makes additional changes
in the set $\mathcal{F}_{t_i}$ of fingerprints in $\Gsat{t_i}$
based on the outcomes of the re-identification, specifically on the level
of uncertainty suffered because of $\mathcal{F}_{t_{i-1}}$.
To that end, she selects a subset $Y'_{t_{i-1}}$ of victims whose fingerprints
were the least useful during the re-identification attempt, in the sense
that they were the most likely to lead to a larger number of equally likely
options after fingerprint mapping. The adversary modifies the fingerprint
of every $y\in Y'_{t_{i-1}}$ by randomly flipping one edge of the form $(y,s)$,
$s\in S_{t_{i-1}}$, checking that the new fingerprint does not coincide
with a previously existing fingerprint. 

The set $Y'_{t_{i-1}}$ is obtained as follows.
For every victim $y_j\in Y_{t_{i-1}}$, let $p_{j}:R_j\rightarrow[0,1]$
be a probability distribution where
$$R_j=\left\{v\in V_{\Grat{t_{i-1}}}\mid \exists_{X\in\mathcal{X}_{t_{i-1}}}
\exists_{\phi\in\mathcal{Y}_X}\ \phi(v)=y_j\right\}$$ is the set of vertices
mapped to $y_j$ according to some $X\in\mathcal{X}_{t_{i-1}}$
and the corresponding $\mathcal{Y}_X$. That is, for every element
$v\in R_j$, $p_{j}(v)$ represents the probability that $v$ has been mapped
to $y_j$ in the previous re-identification attempt according to some sybil
subgraph candidate and some of the resulting fingerprint matchings,
and is defined~as
\begin{displaymath}
p_j(v)=\frac{\sum_{X\in\mathcal{X}_{t_{i-1}}}
\frac{\sum_{\phi\in\mathcal{Y}_X}
\tau_{j}(\phi,v)}{|\mathcal{Y}_X|}}{|\mathcal{X}_{t_{i-1}}|}
\end{displaymath}
where
\begin{displaymath}
\tau_j(\phi,v)=\left\{
\begin{array}{ll}
1&\text{ if }v\in\dom(\phi)\text{ and }\phi(v)=y_j\\
0&\text{ otherwise.}
\end{array}
\right.
\end{displaymath}

Finally, the set $Y'_{t_{i-1}}$ is obtained by making
$$Y'_{t_{i-1}}=\argmax_{y_j\in Y_{t_{i-1}}} \{H(p_j)\},$$
where $H(p_j)$ is the entropy of the distribution $p_j$, that is
$$H(p_j)=-\sum_{v\in R_j} p_j(v)\log p_j(v).$$
If the maximum entropy value is reached for more than one victim,
all of their fingerprints are modified. We chose to use entropy
for obtaining $Y'_{t_{i-1}}$ because it is a well established quantifier
of uncertainty.

\subsubsection*{Targeting new victims}
\label{sssec:attack-new-victims}

After the fingerprints of existing victims have been updated,
the final step of sybil subgraph update consists in targeting
new victims. To that end, the adversary chooses a set of vertices
$y_1,y_2,\ldots,y_r\in V_{t_i}\setminus Y_{t_{i-1}}$,
with $0\le r\le \min\left\{|V_{\Grat{t_{i-1}}}|-|Y_{t_{i-1}}|,
2^{|S_{t_{i-1}}|}-|Y_{t_{i-1}}|-1 \right\}$ and adds them to $Y_{t_i}$.
The new victims can either be fresh vertices, that is vertices that first appear
in $\Gsat{t_i}$, or previously untargeted ones.
The fingerprints of the new victims are created in the same manner
as those of the initial victims. That is, for every $y_j$, $1\le j\le r$,
the vertex $(y_j,s)$, $s\in S_{t_i}$, is added to $\Gsat{t_i}$
with probability $0.5$, checking that the newly generated fingerprint
is different from those of all previously targeted victims.

%===================================================
\section{Experimental Evaluation}
\label{sec:experiments}
%===================================================

In this section, we experimentally evaluate our new dynamic active attack.
Our evaluation has three goals. First, we show that our attack outperforms
Mauw \emph{et al.}'s static robust active attack~\cite{MRT19} in terms
of both effectiveness and efficiency. For simplicity, throughout this section
we will use the acronym D-AA for our attack and S-RAA
for the static robust active attack. Secondly, we determine the factors
that affect the performance of our new attack, and evaluate their impact.
From this analysis, we derive a number of recommendations allowing
data owners to balance privacy preservation and utility
in random perturbation methods for periodical social graph publication.
Due to the scarcity of real-life temporally labelled social graphs,
we run the aforementioned experiments on synthetic dynamic social graphs.
We developed a synthesiser that can be flexibly configured to generate
synthetic dynamic social graphs with specific properties,
e.g. the initial number of nodes and the growth rate. Using this synthesiser,
we run each instance of our experiments in a collection of $100$ synthetic
datasets, which allows to mitigate the impact of random components of our attack
and the synthesiser itself. To conclude, we replicate some of the previous
experiments on a real-life dataset, to show that some of the findings obtained
on synthetic data remain valid in practical scenarios.

%---------------------------------------------------
\subsection{Experimental Setting}
%---------------------------------------------------

We implemented an evaluation tool based on the model described
in Sect.~\ref{sec:framework}. A dynamic social graph simulator loads
a real-life dataset, or uses the synthesiser, to generate the sequence
$\mathcal{G}=(\Gat{t_1}, \Gat{t_2}, \ldots, \Gat{t_i}, \ldots)$
containing only legitimate users. Each snapshot is then processed
by a second module that simulates sybil subgraph creation or update.
The output, which is the data owner's view of the social graph, is processed
by graph perturbation module. In this module, we implement a simple
perturbation method consisting in the addition of cumulative noise.
Finally, a fourth module simulates the re-identification on the perturbed
graph and computes the success probability of the attack.
Sybil subgraph creation and update,
as well as re-identification, are discussed in detail in Sect.~\ref{sec:attack}.
We describe in what follows the implementation of the remaining modules.

%---------------------------------------------------
\subsubsection{Dynamic social graph simulator}
\label{sssec:simulator}
%---------------------------------------------------

Our simulator allows us to conduct experiments on temporally annotated real-life
datasets, as well as synthetic datasets. In the first case, the simulator
extracts from the dataset the graph snapshots by using a specific handler.
The simulator is parameterised with a sequence of the time-stamps
indicating when each snapshot should be taken. Every snapshot is built
by taking all vertices and edges created at a moment earlier or identical
to the corresponding time-stamp and still not eliminated. 

As we mentioned, 
we include in our simulator a synthesiser for generating periodically released
dynamic social graphs, which is based on the Barab\'{a}si-Albert (BA) 
generative graph model~\cite{AB02}. We use BA because it preserves 
the properties of real social graphs, namely power-law degree 
distribution~\cite{AB02}, shrinking diameter~\cite{LKF07},
and preferential attachment. 
The BA model generates scale-free networks by iteratively adding vertices
and creating connections for the newly added vertices using a preferential
attachment scheme. This means that the newly added vertices are more likely
to be connected to previously existing nodes with larger degrees.
The BA model has two parameters: the number of nodes $n_0$
of a (small) seed graph, and the initial degree $M_e$ ($M_e\le n_0$)
of every newly added node. The initial seed graph can be any graph.
In our case we use a complete graph $K_{n_0}$. Every time a new node $v$
is added to the current version $G$ of the BA graph, $M_e$ edges are added
between $v$ and randomly selected vertices in $V_G$. The probability
of selecting a vertex $w\in V_G$ for creating the new edge $(v,w)$
is $\frac{\delta_G(w)}{\sum_{x\in V_G}\delta_G(x)}$, as prescribed
by preferential attachment.

To generate the graph sequence, the synthesiser takes four parameters as input:
\begin{itemize}
\item The parameter $n_0$ of the BA model.
\item The parameter $M_e$ of the BA model.
\item The number of vertices of the first snapshot $n_v$.
\item The growth rate $r_\Delta$, defined as the proportion of new edges compared
to the previous number.
\end{itemize}
The parameters $n_v$ and  $r_\Delta$ determine when snapshots are taken.
The first snapshot is taken when the number of vertices of the graph generated
by the BA model reaches $n_v$, and every other snapshot is taken
when the ratio between the number of new edges
and that of the previous snapshot reaches $r_\Delta$.

%---------------------------------------------------
\subsubsection{Graph perturbation via cumulative noise addition}
\label{sssec:attack-noise-addition}
%---------------------------------------------------

To the best of our knowledge, all existing anonymisation methods against
active attacks based on formal privacy properties \cite{MRT18,MRT18b}
assume a single release scenario, and are thus insufficient
for handling multiple releases. Proposing formal privacy properties
that take into account the specificities of the multiple release scenario
is part of the future work. In our experiments, 
we adapted the other known family of perturbation methods,
random noise addition, to the multiple release scenario.

To account for the incrementality of the publication process, the noise is added
in a cumulative manner. That~is, when releasing $\Grat{t_i}$, the noise
incrementally added on $\Grat{t_1}$, $\Grat{t_2}$, \ldots, $\Grat{t_{i-1}}$
is re-applied on the pseudonymised graph $\varphi_{t_i}\Gsat{t_i}$
to obtain an intermediate noisy graph $\GratInit{t_i}$, and then fresh noise
is added on $\GratInit{t_i}$ to obtain the graph $\Grat{t_i}$ that is released.
In re-applying the old noise, all noisy edges incident in a vertex
$v\in V_{\Grat{t_{i-1}}}\setminus V_{\varphi_{t_i}\Gsat{t_i}}$,
removed after the release of $\Grat{t_{i-1}}$, are forgotten.
The fresh noise addition consists in randomly flipping
a number of edges of $\GratInit{t_i}$. For every flip,
a pair $(v,w)\in V_{\GratInit{t_i}}\times V_{\GratInit{t_i}}$ is uniformly
selected and, if $(v,w)\in E_{\GratInit{t_i}}$, the edge is removed,
otherwise it is added.
The cumulative noise addition method has one parameter: the amount
of fresh noise to add in each snapshot, called \emph{noise ratio}
and denoted $\Omega_{\it noise}$. It~is computed
with respect to $|E_{\GratInit{t_i}}|$, the number of edges
of the pseudonymised graph after restoring the accumulated noise.

\subsubsection{Success probability}

As in previous works on active attacks for the single release
scenario~\cite{MRT18,MRT18b,MRT19}, we evaluate the adversary's success
in terms of the probability that she correctly re-identifies
all victims, which in our scenario is computed by the following formula
for the $t_i$-th snapshot:
\begin{displaymath}
Pr_{succ}^{(t_i)}=\left\{
\begin{array}{ll}
\frac{\sum_{X\in\mathcal{X}_{t_i}}p_X^{(t_i)}}{|\mathcal{X}_{t_i}|}
&\text{ if }\mathcal{X}_{t_i}\ne\emptyset\\
0&\text{ otherwise}
\end{array}
\right.
\end{displaymath}
where
\begin{displaymath}
p_X^{(t_i)}=\left\{
\begin{array}{ll}
\frac{1}{|\mathcal{Y}_{X}|}
&\text{ if }\exists_{\phi\in\mathcal{Y}_{X}}\ \phi^{-1}=\varphi_{t_i}|_{Y_{t_i}}\\
0&\text{ otherwise}
\end{array}
\right.
\end{displaymath}
and, as discussed in Sect.~\ref{ssec:game-components}, $\varphi_{t_i}$
is the isomorphism applied on $\Gsat{t_i}$ to obtain
the pseudo\-nymised graph $\varphi_{t_i}\Gsat{t_i}$.
For every snapshot $\Grat{t_i}$, we compute success probability after
the re-identification refinement is executed.

\begin{figure*}[t]
\centering
\includegraphics[scale= 0.26]{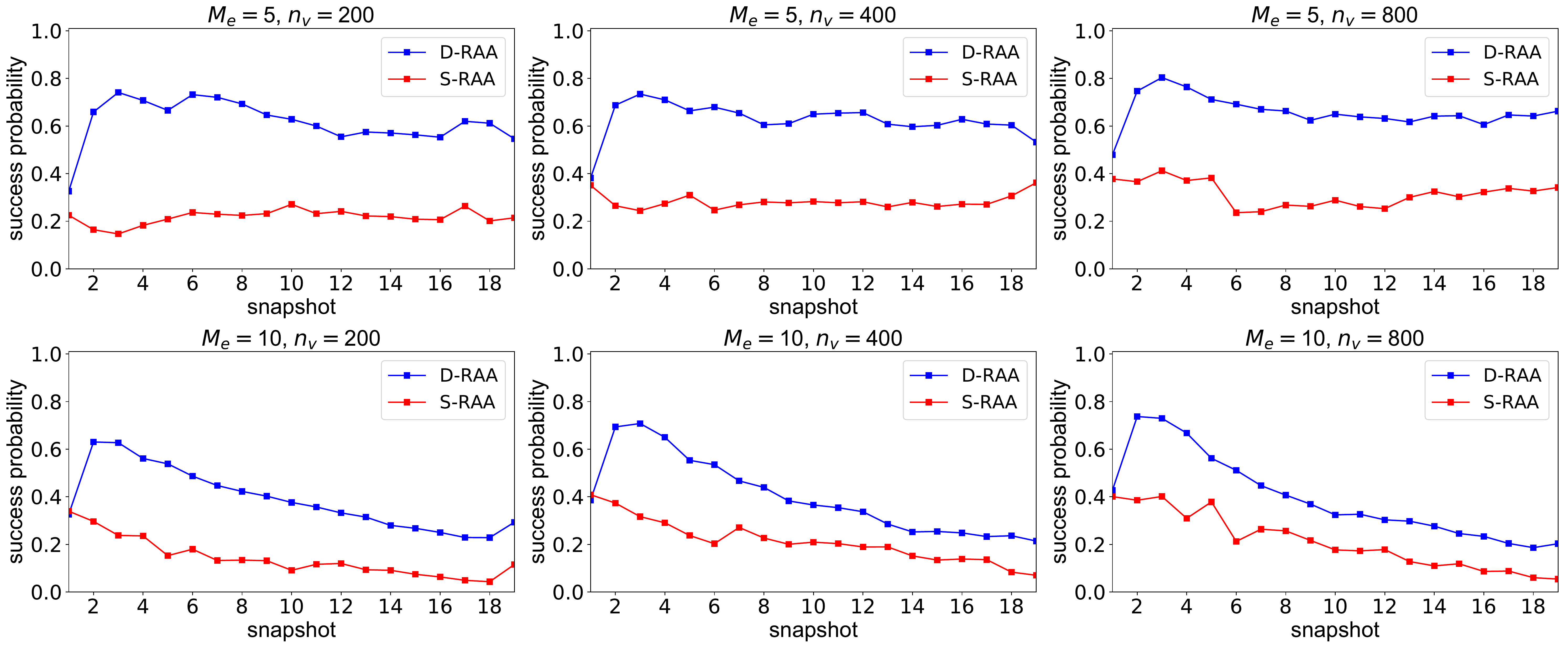}
\caption{Effectiveness comparison between {S-RAA} and {D-AA}.
\label{fig:performance}}
\end{figure*}

%---------------------------------------------------
\subsection{Results and Discussion}
\label{ssec:discussion}
%---------------------------------------------------

We begin our discussion with the comparison of \mbox{D-AA} and {S-RAA}.
Then, we proceed to study the factors that affect the effectiveness of our attack,
and characterise their influence.
Finally, we use the real-life dataset Petster~\cite{K13} to illustrate
the effectiveness of our attack in practice.
For the first two sets of results, we use synthetic dynamic graphs generated
by our synthesiser. Table~\ref{tab:paras} summarises the different configurations
used for the generation. Furthermore, before each release, we select
at least 1 and maximum 5 random legitimate vertices as new victims.
For each parameter combination, we generated $100$
synthetic dynamic graphs, and the results shown are the averages over each
subcollection.

\begin{table}[h]
\centering
\begin{tabular}{|l|l|l|l|l|l|}
\hline
\textbf{Sect.} & ${\bf n_0}$ & ${\bf M_e}$ & ${\bf n_v}$ &
${\bf r_\Delta}$ & ${\bf \Omega_{\it noise} (\%)}$ \\ \hline
\hline
\textbf{5.2.1} & 30 & 5, 10 & 200, 400,800 & 5\% & 0.5 \\ \hline
\textbf{5.2.2} & 30 & 5, 10 & 2000, 4000, 8000 & 5\% & 0.5, 1.0, 1.5, 2.0 \\ \hline
\end{tabular}
\caption{Combinations of parameters for the simulator.\label{tab:paras}}
\end{table}

%----------------------------------------------------
\subsubsection{Comparing {D-AA} and {S-RAA}}
%----------------------------------------------------

The goal of this comparison is to show that our dynamic active attack outperforms
the original attack in both effectiveness and efficiency.
We use six settings for the dynamic graph synthesiser. For each value of $M_e$ 
($5$~and~$10$), we set the initial number of vertices $n_v$ at $200$, $400$ 
and $800$. In all our experiments, sybil subgraph creation spans the first 
and second snapshots, and the re-identification is executed for the first time 
on the second snapshot. 

\smallskip
\noindent{\bf Effectiveness comparison.}
In Fig.~\ref{fig:performance} we show the success probabilities of the two
attacks on graphs with different initial sizes and
amounts of changes between consecutive releases (determined by $M_e$ with
$r_\Delta$ fixed as $5\%$).

We have three major observations from the results.
First, we can see that
{D-AA} significantly outperforms {S-RAA} in terms of success probability.
The improvement becomes larger when smaller numbers of changes occur between
releases. {D-AA} outperforms {S-RAA} by at least twice,
even up to three times for the first few snapshots.
When graphs grow slowly ($M_e=5$),
our attack always displays an average success probability larger than 0.5
which {S-RAA} never reaches.
Second, the success probability of the original {S-RAA} has a general trend
to drop along with time, while {D-AA} displays a large increase
from the first to the second snapshot, and then remains stable
or degrades very slowly.
This is the result of the reduced uncertainty enabled by the temporal consistency
constraints. The frequent modification of the set of sybil nodes allows our attack
to offset the noises accumulated over time and maintain
an acceptable success probability even at the later releases.
When $M_e$ is set to 10, the success probability remains above 0.5 until
the sixth snapshot, even though noise grows faster in this case.
Third, {S-RAA} is more likely to be
influenced by the randomness of graph structures and noise,
as shown by the large fluctuations
of the success probability, while our {D-AA} displays smaller
variance and smoother curves.

\begin{figure*}[!t]
\centering
\includegraphics[scale= 0.26]{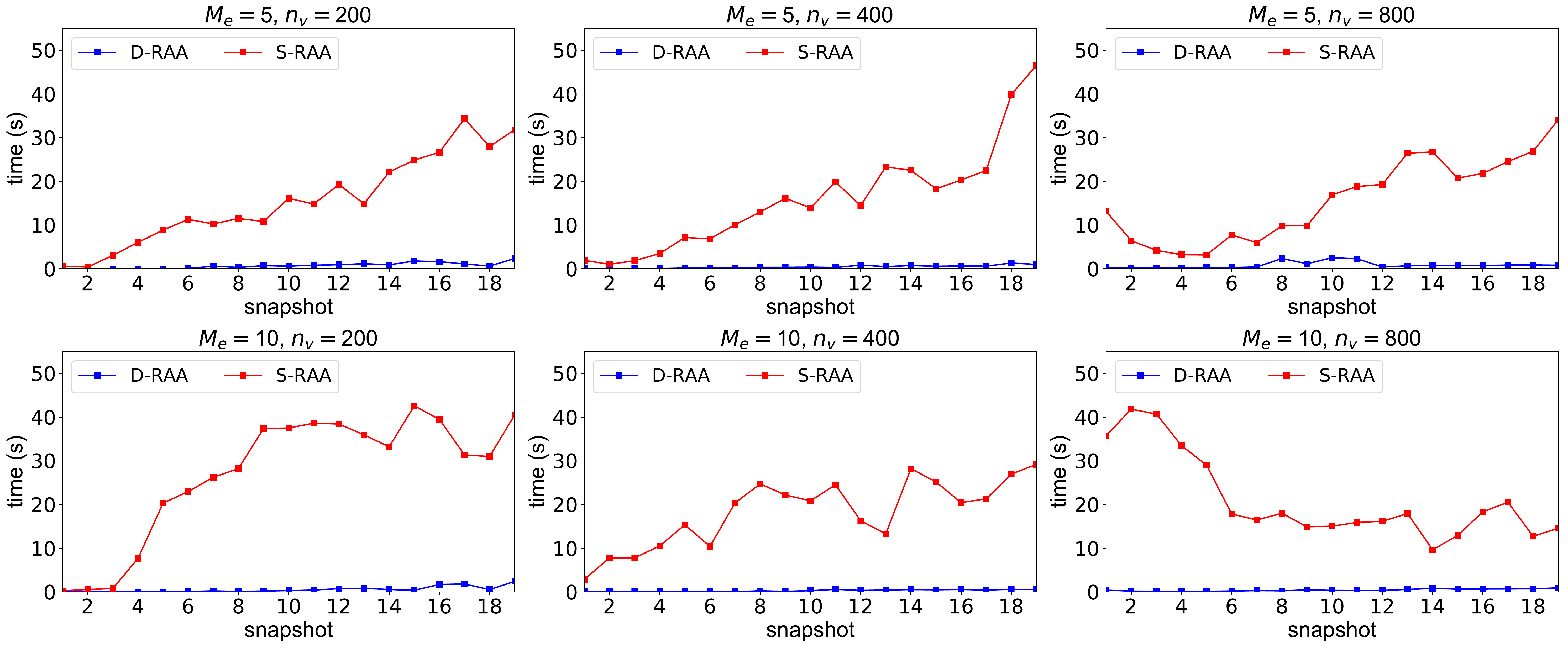}
\caption{Efficiency comparison between {S-RAA} and {D-AA}.
\label{fig:efficiency}}
\end{figure*}

\smallskip
\noindent{\bf Efficiency comparison.}
Fig.~\ref{fig:efficiency} shows the average amount of time consumed
by {S-RAA} and {D-AA} in different scenarios.
We can see that {D-AA} takes an almost constant amount of time at all snapshots,
whereas the time consumption of {S-RAA} grows considerably along time.
This clearly shows that the use of temporal information in dynamic social graphs
helps {D-AA} to effectively avoid the computation overhead.
We highlight the fact that \mbox{D-AA} runs at least 10 times faster
in almost all cases, especially in late snapshots.
An interesting observation is that the running time of {S-RAA} decreases
when $M_e=10$ and the size of the initial graph is 800. Rather than an improvement,
this is in fact the consequence of the repeated failure
of the sybil subgraph retrieval algorithm to find any candidates.
This problem is in turn caused by the small tolerance threshold
required by {S-RAA} to complete runs in reasonable time.

\subsubsection{Factors influencing our attack}

We intend this analysis to serve as a guide for customising the settings
of privacy-preserving publication methods for dynamic social graphs,
in particular for determining the amount of perturbation needed to balance
the privacy requirements and the utility of published graphs.
Since our attack is considerably efficient, we will use for this evaluation
dynamic graphs featuring 2000, 4000 and 8000 initial vertices.
In addition, we set large tolerance thresholds 
for structural dissimilarity in the sybil subgraph retrieval method, 
which increases the probability that it finds $\varphi_{t_i}(S_{t_i})$ 
as a candidate. 
In our experiment, 
we set the threshold $\theta_{t_i}$ used at the \mbox{$i$-th} snapshot 
to $\theta_{t_i}=\min(1500, 16+250\times(i-2)^2)$. 
Three factors may possibly
impact the effectiveness of re-identification attacks on dynamic graphs:
the amount of noise, the size of graphs and the speed of growth between
two releases. As a result, we analyse three parameters which
determine these three factors in our simulator: $\Omega_{\it noise}$,
$M_e$ and $n_v$. The number of vertices in the initial snapshots
$n_v$ determines the scale of the released graphs, while the parameter $M_e$
of the BA model controls the number of new nodes and edges added before
the next release. The noise ratio $\Omega_{\it noise}$ determines
the amount of noise.

\begin{figure*}[!ht]
\centering
\includegraphics[scale= 0.26]{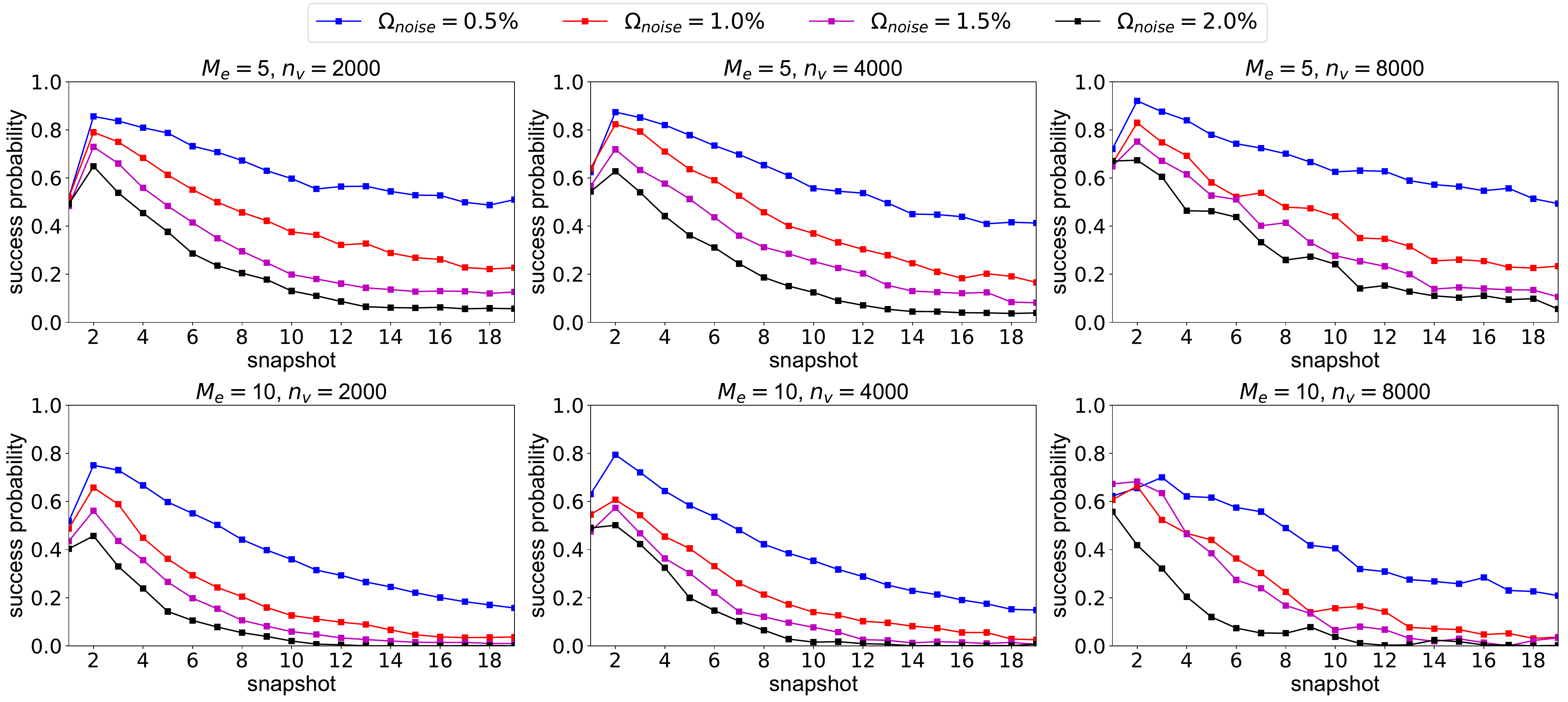}
\caption{Factors influencing the success probability.\label{fig:riskImp}}
\end{figure*}

\begin{figure*}[!ht]
\centering
\includegraphics[scale= 0.26]{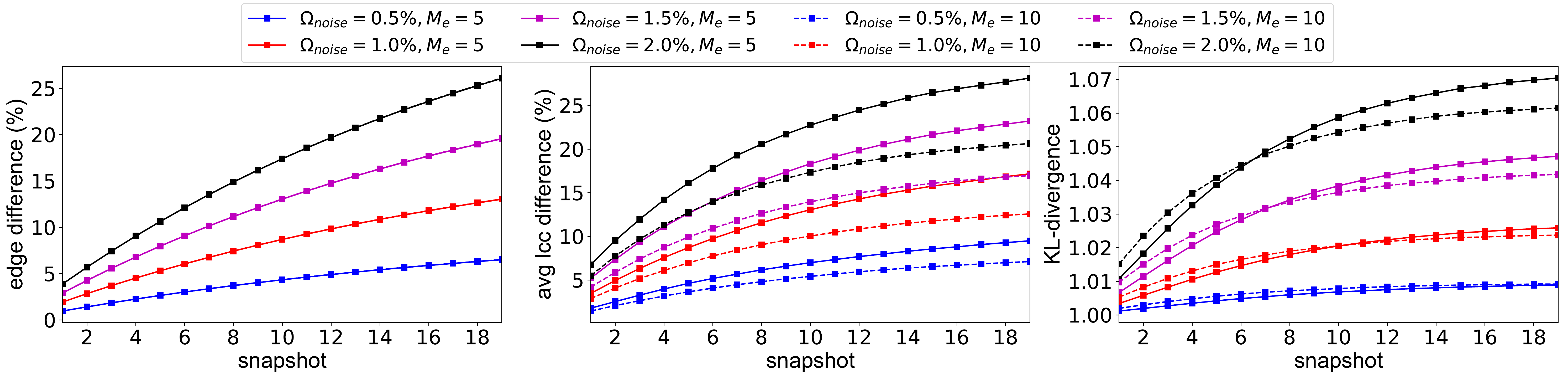}
\caption{Factors influencing the utility of released graphs.\label{fig:usability}}
\end{figure*}

\smallskip\noindent
{\bf Effectiveness.}
Fig.~\ref{fig:riskImp} shows the success probability of our attack
when different noise ratios are applied on dynamic graphs with different
initial sizes and growth speeds.
First, we can see that the success probability decreases when more noise is
applied. This is natural as more perturbation makes it more difficult to find
the correct sybil sugraph, either because the sybil graph has been too
perturbed to be found as a candidate
or because edge perturbation generates more subgraphs similar
to the original sybil subgraph.
When $M_e=5$ and the noise ratio is set to $0.5\%$, success probability
always remains above $0.5$. For this value of $M_e$,
even with $\Omega_{\it noise}$ at $2.0\%$, the attack still displays
success probability above $0.5$ in the first three snapshots.
Second, the rate at which success probability decreases slows down
as we increase the value of the noise ratio.
The largest drop occurs when we increase $\Omega_{\it noise}$ from $0.5$ to $1.0$.
This suggests that keeping increasing the level of perturbation may not
necessarily guarantee a better privacy protection, but just damage the utility
of the released graphs.
Third, the sequences of success probability values show a very small dependence
on the initial size of the graphs, with other parameters fixed.
Last, the success probability decreases when dynamic graphs grow faster.
From the figure, we can see that the probability reduces by about $10\%$ when
we increase the value of $M_e$ from 5 to 10.

Summing up, we observe that the risk of re-identification decreases
when more perturbation is applied and when the graphs grow faster,
whereas the initial size of the graphs has a relative small impact on this risk.

\smallskip\noindent
{\bf Utility.}
We evaluate the utility of released graphs in terms of three measures: 
the percentage
of edge editions, the variation of the average local clustering coefficient,
and the KL-divergence of degree distributions.
The first measure quantifies the percentage of edge flips with respect to the
total number of edges. In fact, it quantifies the amount of noise accumulated so
far. For the $t_i$-snapshot, the percentage of edge editions is computed~as
\[
\frac{1}{\mmid E_\Gsat{t_i}\mid}
\mmid\set{(v,v')\in E_\Gsat{t_i}\mid (\varphi(v),
\varphi(v'))\not\in\Grat{t_i}} \\
\cup\set{(v,v')\in E_\Grat{t_i}\mid(\varphi^{-1}(v),
\varphi^{-1}(v'))\not\in\Gsat{t_i}}\mmid.
\]

The local clustering coefficient (LCC) of a vertex measures
the proportion of pairs of mutual neighbours of the vertex that are connected by
an edge. We calculate the average to the LCCs of all vertices and take the
proportion between the differences of the value of the original graph and that
of the anonymised graph as, that is 
\[
\frac{\mmid {\it avgLcc}(\Gsat{t_i})-{\it avgLcc}(\Grat{t_i})\mmid}
{{\it avgLcc}(\Gsat{t_i})}.
\]
where ${\it avgLcc}(G)$ is the average local clustering coefficient
of graph $G$.
Our last measure uses the KL-divergence~\cite{KL51} as an indicator
of the difference between the degree distribution of the original graph
and that of the perturbed graph.

As all the three measures present almost identical patterns for different
values of $n_v$, we only show the results for $n_v=8000$.
We have two major observations. First, as expected, the values of all three measures
increase as the noise accumulates over time, indicating that the utility of released
graphs deteriorates. Even with $\Omega_{\it noise}$ set to just $1.0\%$,
at the tenth snapshot we can have up to $10\%$ of edges flipped and changes
in edge density around $15\%$. At this point, we can say that the utility
of released graphs has already been greatly damaged.
Second, when dynamic graphs grow faster, the impact of noise becomes smaller,
as more legitimate edges offset the impact of noisy edges.

Together with the finding that larger growth
speed results in smaller success probability, we can conclude
that the social networks that grow fast among releases display a better balance
between re-identification risk and the utility of the released graphs.

\subsubsection{Results on a real-life dynamic social graph}

We make use of a publicly available graph collected from Petster, a website for
pet owners to communicate~\cite{K13}, to validate to what extent the results
reported in the previous subsection remain valid in a more realistic domain.
The Petster dataset is an undirected graph whose vertices
represent the pet owners. The vertices are labelled by their joining date, which
span from January 2004 to December 2012. The graph is incremental, which means
no vertices are removed. It contains 1898 vertices and $16,750$ edges.
We take a snapshot every six months.

\begin{figure}[h]
\centering
\subfigure[Re-identification risk on Petster.\label{fig:riskPetster}]
{\includegraphics[scale= 0.27]{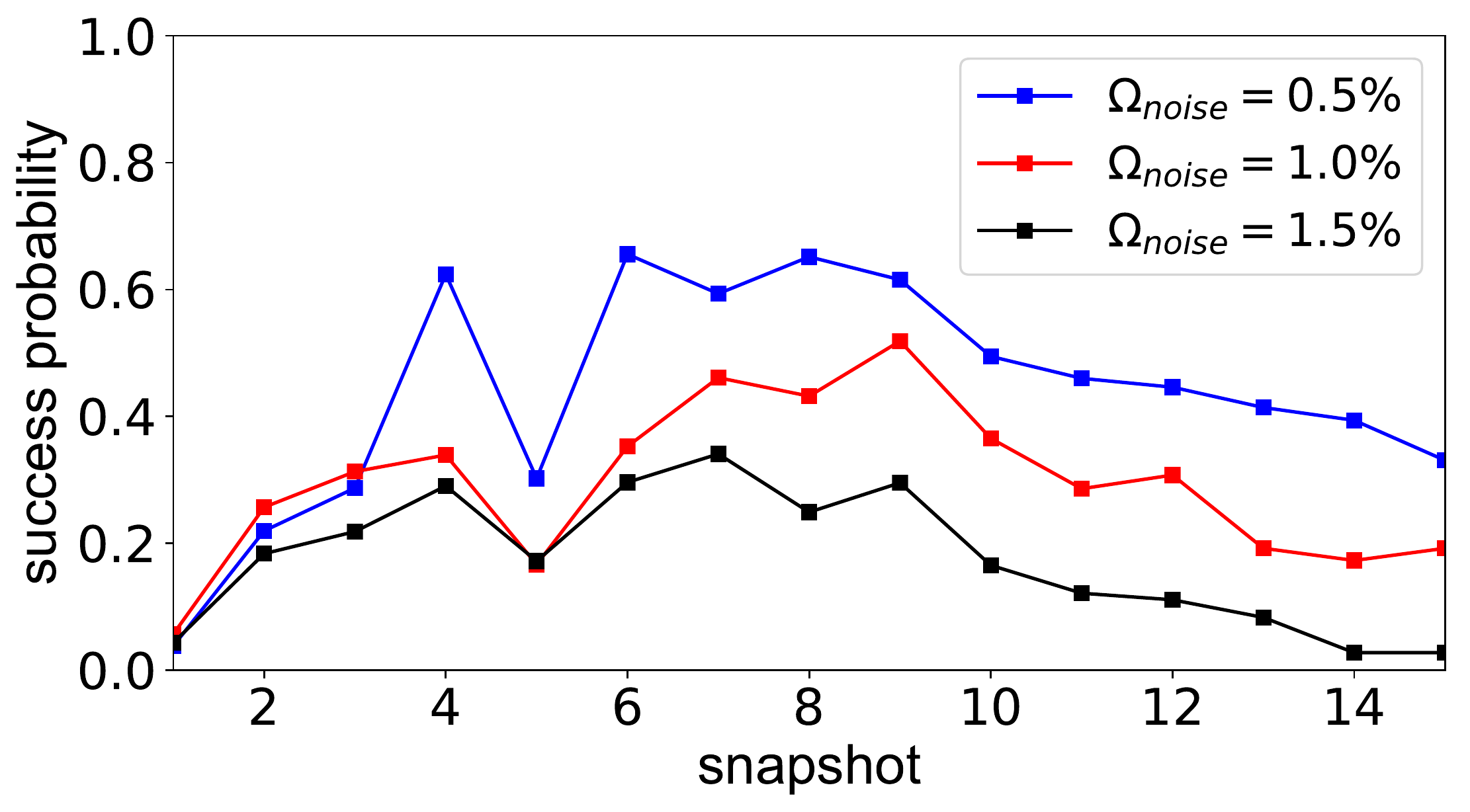}}
\subfigure[Numbers of new vertices added before release.\label{fig:newV}]
{\includegraphics[scale= 0.27]{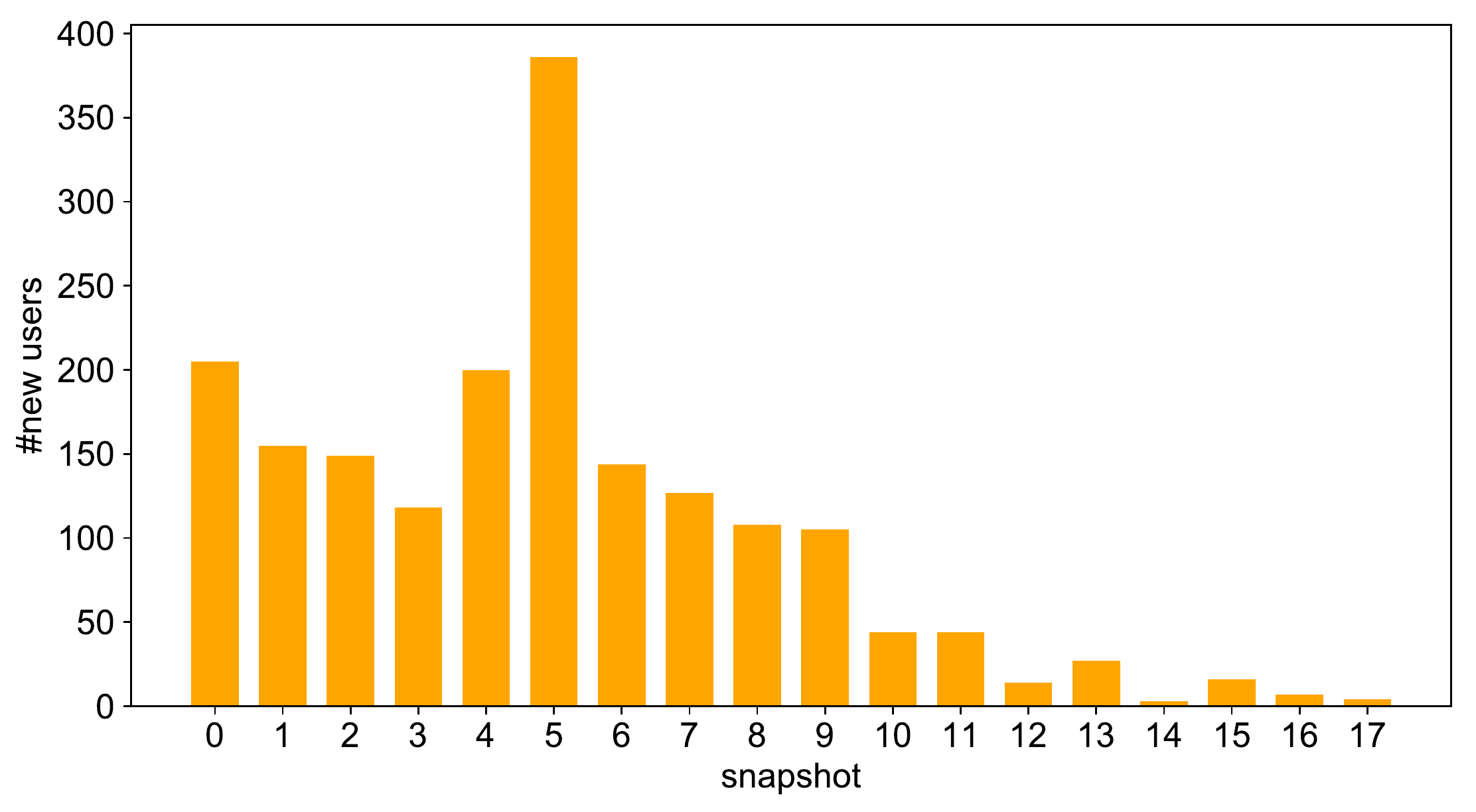}}
\caption{Evaluation on Petster.}
\end{figure}

We present in Fig.~\ref{fig:riskPetster} the success probability of our {D-AA}
attack on the Petster dataset when the noise ratio is set to $0.5\%$, $1.0\%$
and $1.5\%$, respectively.
Compared to the success probabilities discussed above on
simulated graphs, the curves have different shapes and more fluctuations.
This is because, instead of a fixed growth speed (determined by $r_\Delta$ and
$M_e$ in our simulator), the
real-life graph grows at different speeds in different periods, as shown in
Fig.~\ref{fig:newV}. We can see that the number of new vertices varies before
each release. After the first few years of steady growth, Petster gradually lost
its popularity, especially with few new vertices added in the last three years.
By cross-checking the two figures, we can see that the success probability
changes with the amount
of growth before the corresponding release. It first increases steadily due to
the steady growth of the graph until the fifth snapshot, which suddenly has the
largest number of new vertices. Then when the growth slows down, the success
probability also
recovers; and when the growth stops (e.g., from the 12 snapshot), it starts
increasing again, even though the noise continues to accumulate.
These observations validate our findings
on synthetic graphs, that is, the speed of growth is the dominating factor
that affects the re-identification risk.

%===================================================
\section{Conclusions}
\label{sec:conclusion}
%===================================================

In this paper, we have presented the first dynamic active re-identification 
attack on periodically released social graphs. Unlike preceding attacks, 
the new attack exploits the inherent dynamic nature of social graphs 
by leveraging tempo-structural patterns for re-identification. 
Compared to existing (static) active attacks, our new dynamic attack 
significantly improves success probability, by more than two times, 
and efficiency, by almost 10 times. Through comprehensive experimental 
evaluation on synthetic data, we analysed the factors influencing the success
probability of our attack, namely the growth rate of the graph 
and the amount of noise injected. These findings can subsequently be used 
to develop graph anonymisation methods that better balance privacy protection 
and the utility of the released graphs. 
For instance, for a given
noise level, the decision to publish a new snapshot should be determined
by taking into
account the number of changes that have occurred from the last release. Similarly,
if the time for the next release is given, the amount of noise should be
customised according to the number of changes that have occurred. 
Additionally, we evaluated our attack on Petster, a real-life dataset. 
This evaluation showed that some of the findings obtained
on synthetic data remain valid in practical scenarios.

\vspace*{.7cm}
\noindent \textbf{Acknowledgements:} The work reported in this paper 
received funding from Luxembourg's Fonds National de la Recherche (FNR), 
via grant C17/IS/11685812 (PrivDA).

\end{document}